\documentclass[10pt]{article}
\usepackage{graphicx}
\usepackage{amsmath}
\usepackage{amssymb}
\usepackage{caption2}
\begin{document}
\bibliographystyle{prsty}
\begin{center}
{\large {\bf \sc{  Two-body strong decays of the $P_{cs}(4338)$ as a hidden-charm pentaquark  state  via the QCD sum rules }}} \\[2mm]
Zhi-Gang Wang \footnote{E-mail: zgwang@aliyun.com.  }, Jia-Wei Zhu, Yang Liu     \\
 Department of Physics, North China Electric Power University, Baoding 071003, P. R. China
\end{center}

\begin{abstract}
In this work, we study the hadronic coupling constants of the $P_{cs}(4338)$ (listed as $P_{c\bar{c}s}(4338)^0$  by the Particle Data Group) as a  $[us][dc]\bar{c}-[ds][uc]\bar{c}$-type pentaquark state with  the spin-parity $J^P={\frac{1}{2}}^-$  via the three-point QCD sum rules according to rigorous  quark-hadron duality for the first time. We select the optimal parametrization to subtract  contaminations from the higher resonances and continuum states and  obtain reasonable  hadronic coupling constants therefore  the partial decay widths and total width  $\Gamma(P_{cs}(4338))=10.48\pm 2.41\,\rm{MeV}$ ($8.90\pm 2.03\,\rm{MeV}$), which are consistent  with the experimental data  $\Gamma(P_{cs}(4338)) = 7.0\pm1.2\pm1.3 \mbox{ MeV}$ from the LHCb collaboration. And we can compare the partial decay widths  to the precise experimental data in the future.
\end{abstract}



\section{Introduction}
In 2023,  the  LHCb collaboration observed a narrow resonance in the $J/\psi\Lambda$ system, which is consistent with a pentaquark candidate with strangeness \cite{LHCb-Pcs4338-2023}. The mass and  width  are $4338.2\pm 0.7\pm 0.4\, \rm{MeV}$ and ${7.0\pm1.2\pm1.3 \, \rm{MeV}}$ respectively, and preferred spin-parity is $J^P={\frac{1}{2}}^-$. The thresholds of the $\bar{D}\Xi_c$, $\bar{D}^*\Xi_c$, $\bar{D}\Xi^\prime_c$, $\bar{D}_s\Lambda_c$ and $\bar{D}^*_s\Lambda_c$ are $4337\,\rm{MeV}$, $4478\,\rm{MeV}$, $4446\,\rm{MeV}$, $4255\,\rm{MeV}$ and $4399\,\rm{MeV}$ respectively from the Particle Data Group \cite{PDG}, which stimulates the assignment of the $P_{cs}(4338)$ as the $\bar{D}\Xi_c$ molecular state \cite{WangZG-review,Pcc4338-mole-WangFL-PLB-2022,Pcc4338-mole-Feijoo-PLB-2023,
Pcc4338-mole-Ozdem-PLB-2023,Pcc4338-mole-MengL-PRD-2023,
Pcc4338-mole-Ortega-PLB-2023,Pcc4338-mole-YanMJ-PRD-2023,
 Pcc4338-mole-Nakamura-PRD-2023,Pcc4338-mole-HeJ-PRD-2023,
Pcc4338-mole-LiHS-PRD-2024,Pcc4338-mole-LiHS-PRD-2024-2,Pcc4338-mole-WangB-PRD-2024,
Pcc4338-mole-PingJL-Symmetry-2024,Pcs4338-mole-WangXW-CPC-2023}.

It is obvious that we cannot identify a hadron unambiguously based on the mass alone, we have to study the strong decays, weak decays, electromagnetic decays, productions so as to reach a more robust identification \cite{Pcs4338-mole-decay-WangXW-PRD,
Pcs4338-mole-decay-Azizi-PRD-2023,Pcs4338-penta-MM-Ozdem-PRD-2026,
Pcs4338-Braaten-BO,
Pcc4338-Product-Clymton-PRD-2025,Pcs4338-Product-HeJ-EPJC-2025,
Pcs4338-Product-HsiaoYK-PRD-2025,Pcs4338-Product-ChenDY-PRD-2024}.

In Ref.\cite{Pcs4338-mole-WangXW-CPC-2023},  we  assigned  the $P_{cs}(4338)$ as the $\bar{D}\Xi_c$  molecular state tentatively, and distinguished  isospins of the currents to  study  the relevant  molecular states  without strange, with strange  and with double-strange  via  the QCD sum rules systematically.  In Ref.\cite{Pcs4338-mole-decay-WangXW-PRD}, we studied  the two-body strong decays of the $P_{cs}(4338)$ and its higher isospin cousin $P_{cs}(4460)$ (which is still not observed yet) as molecular states  via the three-point QCD sum rules, and obtained the partial decay widths,
$\Gamma ( \eta_c \Lambda) = 0.95^{+0.50}_{-0.50} \,{\rm MeV}$ and
$\Gamma ( J/\psi \Lambda) = 5.21^{+8.00}_{-5.21} \,{\rm MeV}$.
In Ref.\cite{Pcs4338-mole-decay-Azizi-PRD-2023},
 Azizi, Sarac and  Sundu also studied the  decay channels $P_{cs}(4338) \rightarrow J/\psi \Lambda$ and $ \eta_c \Lambda$ via the three-point QCD sum rules, and obtained the partial decay widths $\Gamma(J/\psi \Lambda)=7.22\pm 1.78~\rm{MeV}$ and
 $\Gamma(\eta_c \Lambda)=3.18\pm 0.74~\rm{MeV}$.

In Ref.\cite{Pcs4338-penta-MM-Ozdem-PRD-2026}, Ozdem calculated the magnetic moments of the diquark-diquark-antiquark type hidden-charm pentaquark states with strangeness via the light-cone QCD sum rules.

In Ref.\cite{Pcs4338-Braaten-BO},  Alasiri, Braaten and Bruschini
identified  the  $P_{cs}(4338)$ and $P_{cs}(4459)$ as two members of the ground-state triplet for $c\bar{c}$ in
the potential associated with a spin-$\frac{1}{2}$ strange adjoint baryon
in the Born-Oppenheimer  approximation, and obtained the ratio,
\begin{eqnarray}
\frac{\Gamma\left(P_{cs}(4338)\to \eta_c\Lambda\right)}{\Gamma\left(P_{cs}(4338)\to J/\psi\Lambda\right)}&=&0.43\, .
\end{eqnarray}

The literatures on the two-body strong decays of the $P_{cs}(4338)$ are rare, and  no definite conclusion could be obtained. There have been several possible interpretations of the observed hidden-charm pentaquark candidates, such as threshold cusps, triangle singularities, dynamically  generated resonances, molecular states, diquark-diquark-antiquark type pentaquark states, etc. In the present work, we would like to focus on the method of the QCD sum rules \cite{WangZG-review}.

Based on the predictions of the QCD sum rules, we can reproduce the experimental  masses of the  $P_c(4312)$,  $P_{cs}(4338)$, $P_{c}(4380)$, $P_c(4440)$, $P_c(4457)$ and $P_{cs}(4459)$ both in the scenarios of  pentaquark   and molecular states \cite{WangZG-review}, while we can only reproduce the experimental  mass of the $P_c(4337)$ in the pentaquark scenario \cite{WangZG-penta-Jpsi-p}.

In Refs.\cite{WangZG-penta-Jpsi-p,WangZG-penta-IJMPA-2020,WangZG-penta-Jpsi-Lamba,
WangZG-penta-Jpsi-Xi,WangZG-penta-Jpsi-Sigma,WangZG-penta-Jpsi-XiV,
WangZG-penta-Jpsi-Omega,WangZG-penta-Jpsi-SigmaV}, we performed comprehensive investigations  of the mass spectroscopy of the octet and decuplet    diquark-diquark-antiquark type hidden-charm pentaquark  states without strange, with singly-strange, doubly-strange, triply-strange and  with the spin-parity $J^P={\frac{1}{2}}^-$, ${\frac{3}{2}}^-$ and ${\frac{5}{2}}^-$
via the QCD sum rules. In contrast to schemes adopted in other literatures on the exotic states \cite{QCDSR-Narison-PRD-2007,QCDSR-Nielsen-PRT-2010,
 QCDSR-ChenZhu-PRD-2011,QCDSR-Nielsen-JPG-2019,QCDSR-ChenW-PRD-2013,
 QCDSR-ZhangJR-PLB-2011,QCDSR-ZhangJR-PRD-2013,QCDSR-QiaoCF-NPB-2021,
 QCDSR-QiaoCF-PRD-2020}, we carried  out the operator product expansion (OPE)   up to   the vacuum condensates of dimension $13$ in our unique counting  scheme consistently and chose the optimal energy scales of the QCD spectral densities \cite{WangZG-HuangT-PRD-2014,WangZG-formula-EPJC,Wang1508-EPJC-1}, and assigned all the observed  pentaquark candidates reasonably.

There have been a number of works on the two-body strong decays of the exotic states via the QCD sum rules \cite{Decay-Nielsen-PRD-2013,Decay-WangZG-NPA-2014,
Decay-ChenHX-EPJC-2015,Decay-Azizi-PRD-2016,Duality-WangZG-EPJC-2016,
Duality-WangZG-ZhangJX-EPJC-2018,Decay-Azizi-EPJC-2019,Duality-WangZG-EPJC-2019,
Duality-Penta-mole-WangZG-CPC-2020,Duality-Penta-mole-LiuYL-PRD-2020,
Duality-Penta-WangZG-CPC-2021,Decay-Azizi-JPG-2021,Duality-WangZG-CPC-2022,
Duality-WangZG-CPC-2022-2,Decay-Azizi-EPJC-2023,
Decay-SunH-NPB-2023,Duality-WangZG-PRD-2024,Duality-WangZG-AAPPS-2024,
Duality-YangXS-CPC-2025,Duality-WangXW-CPC-2024,
Duality-WangZG-NPB-2024,Decay-Azizi-NPB-2026,Duality-YangXS-CPC-2026}. In contrast to schemes  adopted in other literatures on  studying  the two-body strong decays of the exotic states with the three-point correlation functions  \cite{Decay-Nielsen-PRD-2013,Decay-WangZG-NPA-2014,
Decay-ChenHX-EPJC-2015,Decay-Azizi-PRD-2016,Decay-Azizi-EPJC-2019,
Decay-Azizi-JPG-2021,Decay-Azizi-EPJC-2023,Decay-SunH-NPB-2023,
Decay-Azizi-NPB-2026},
in Refs.\cite{Duality-WangZG-EPJC-2016,
Duality-WangZG-ZhangJX-EPJC-2018,Duality-WangZG-EPJC-2019}, we introduced a free parameter $C$ to parameterize transitions among the ground state traditional hadrons  and higher resonances (and continuum states) in the exotic channels, and we attained rigorous quark-hadron duality in the three-point QCD sum rules for the first time.
Then the rigorous quark-hadron duality was successfully  applied to study the strong decays of the exotic states \cite{
Duality-Penta-mole-WangZG-CPC-2020,Duality-Penta-mole-LiuYL-PRD-2020,
Duality-Penta-WangZG-CPC-2021,Duality-WangZG-CPC-2022,
Duality-WangZG-CPC-2022-2,
Duality-WangZG-PRD-2024,Duality-WangZG-AAPPS-2024,
Duality-YangXS-CPC-2025,Duality-WangXW-CPC-2024,
Duality-WangZG-NPB-2024,Duality-YangXS-CPC-2026}.

In this  work, we study the hadronic coupling constants of the $P_{cs}(4338)$ in the scenario of the diquark-diquark-antiquark type  pentaquark state via the three-point  QCD sum rules according to the  rigorous  quark-hadron duality and try to select  the optimal form of the subtracting constants $C$, then study the two-body strong decays and diagnose the nature of the $P_{cs}(4338)$.

We  arrange the skeleton as:  in Sect.2,  we acquire the QCD sum rules for the  hadronic  coupling constants of the  $P_{cs}(4338)$;  in Sect.3, we present the numerical results and discussions; and Sect.4 is reserved for our
conclusion.

\section{The hadronic coupling  constants of the $P_{cs}(4338)$}
Firstly, we write down  the three-point correlation functions $\Pi(p,q)$ and $\Pi_{\mu}(p,q)$ routinely,
\begin{eqnarray}
\Pi(p,q)&=&i^2\int d^4xd^4y e^{ip \cdot x}e^{iq \cdot y} \langle0|T\left\{J_M(x)J_B(y)\bar{J}_P(0)\right\}|0\rangle \, , \\
\Pi_\mu(p,q)&=&i^2\int d^4xd^4y e^{ip \cdot x}e^{iq \cdot y} \langle0|T\left\{J^\mu_M(x)J_B(y)\bar{J}_P(0)\right\}|0\rangle \, ,
\end{eqnarray}
where $J_M(x)= J_{\eta_c}(x)$, $J_{\bar{D}_s}(x)$,  $J_B(y)= J_{\Lambda_c}(y)$, $J_{\Lambda}(y)$, $J^\mu_M(x)=J^\mu_{J/\psi}(x)$,
\begin{eqnarray}
 J_{\eta_c}(x)&=& \bar{c}(x)i\gamma_5 c(x)\, ,\nonumber\\
 J_{\bar{D}_s}(x)&=& \bar{c}(x)i\gamma_5 s(x)\, ,\nonumber\\
 J^\mu_{J/\psi}(x)&=& \bar{c}(x) \gamma^\mu c(x)\, ,
\end{eqnarray}
\begin{eqnarray}
J_{\Lambda_c}(y)&=&  \varepsilon^{ijk}  u^T_i(y) C\gamma_5 d_j(y)\,  c_{k}(y) \, ,\nonumber \\
 J_{\Lambda}(y)&=&  \frac{\varepsilon^{ijk}}{\sqrt{2}} \left[ u^T_i(y) C\gamma_\alpha s_j(y)\, \gamma^\alpha\gamma_5 d_{k}(y)-d^T_i(y) C\gamma_\alpha s_j(y)\, \gamma^\alpha\gamma_5 u_{k}(y)\right] \, ,
\end{eqnarray}
 \begin{eqnarray}
  J_P(0)&=&\frac{\varepsilon^{ila} \varepsilon^{ijk}\varepsilon^{lmn}}{\sqrt{2}} \left[u^T_j(0) C\gamma_\alpha s_k(0)d^T_m(0) C\gamma^\alpha c_n(0) -d^T_j(0) C\gamma_\alpha s_k(0)u^T_m(0) C\gamma^\alpha c_n(0) \right] C\bar{c}^{T}_{a}(0) \, , \nonumber\\
\end{eqnarray}
the $a$, $i$, $j$, $\cdots$  are color indices. The currents $J_{\eta_c}(x)$, $J_{\bar{D}_s}(x)$, $J_{J/\psi}^\mu(x)$,
$J_{\Lambda_c}(y)$, $J_{\Lambda}(y)$ and  $ J_P(0)$ interpolate the hadrons $\eta_c$, $\bar{D}_s$,
 $J/\psi$, $\Lambda_c$, $\Lambda$ and  $P_{cs}(4338)$, respectively. We adopt the $[us][dc]\bar{c}-[ds][uc]\bar{c}$ type five-quark current, which interpolates the
lowest hidden-charm pentaquark state with strangeness \cite{WangZG-penta-Jpsi-p}.

 On the hadron  side, we  insert  a complete set  of intermediate  hadron  states with the
same quantum numbers as the currents  into the correlation functions $\Pi(p,q)$ and $\Pi_{\mu}(p,q)$  to obtain the hadronic representation \cite{WangZG-review,QCDSR-Nielsen-JPG-2019,SVZ79,PRT85}, then we isolate all the  ground  state contributions,
\begin{eqnarray}
 \Pi_{P\eta_c\Lambda}(p,q)   & = & \frac{f_{\eta_c}m_{\eta_c}^2\lambda_P\lambda_\Lambda}{2m_c}\frac{-i \left(\!\not\!{q}+m_\Lambda \right)\left( \!\not\!{p}^\prime+m_P \right)}{\left(m_P^2-p^{\prime2}\right)\left(m_{\eta_c}^2-p^{2}\right)
 \left(m_{\Lambda}^2-q^{2}\right)}G_{P\eta_c\Lambda}+\cdots\,  ,
\end{eqnarray}

\begin{eqnarray}
 \Pi_{P\bar{D}_s\Lambda_c}(p,q)   & = & \frac{f_{D_s}m_{D_s}^2\lambda_P\lambda_{\Lambda_c}}{m_c}\frac{-i \left(\!\not\!{q}+m_{\Lambda_c} \right)\left( \!\not\!{p}^\prime+m_P \right)}{\left(m_P^2-p^{\prime2}\right)\left(m_{D_s}^2-p^{2}\right)
 \left(m_{\Lambda_c}^2-q^{2}\right)}G_{P\bar{D}_s\Lambda_c}+\cdots\,  ,
\end{eqnarray}

\begin{eqnarray}
 \Pi^\mu_{PJ/\psi\Lambda}(p,q)  & = & f_{J/\psi}m_{J/\psi}\lambda_P\lambda_\Lambda\frac{- \left(\!\not\!{q}+m_\Lambda \right)\left( G_V\gamma^\alpha-i\frac{G_T}{m_P+m_\Lambda}\sigma^{\alpha\beta}p_\beta\right)
 \gamma_5\left( \!\not\!{p}^\prime+m_P \right)}{\left(m_P^2-p^{\prime2}\right)\left(m_{J/\psi}^2-p^{2}\right)\left(m_{\Lambda}^2-q^{2}\right)}\nonumber\\
  &&\left(-g_{\mu\alpha}+\frac{p_{\mu}p_{\alpha}}{p^2} \right)+\cdots\,  ,
\end{eqnarray}
where $p^\prime=p+q$, we add the subscripts $P\eta_c\Lambda$, $P\bar{D}_s\Lambda_c$ and $PJ/\psi\Lambda$ to distinguish the  hadronic coupling constants, and we  adopt the conventional  definitions for the pole residues,
\begin{eqnarray}
\langle 0| J_P(0)|P_{cs}(p^\prime)\rangle &=&\lambda_{P} U(p^\prime,s) \, ,  \nonumber\\
\langle 0| J_{\Lambda} (0)|\Lambda(q)\rangle &=&\lambda_{\Lambda} U(q,s) \, , \nonumber \\
\langle 0| J_{\Lambda_c} (0)|\Lambda_c(q)\rangle &=&\lambda_{\Lambda_c} U(q,s) \, , \nonumber \\
\langle 0| J_{\eta_c} (0)|\eta_c(p)\rangle &=&\frac{f_{\eta_c}m_{\eta_c}^2}{2m_c} \, ,\nonumber \\
\langle 0| J_{\bar{D}_s} (0)|\bar{D}_s(p)\rangle &=&\frac{f_{D_s}m_{D_s}^2}{m_c} \, ,\nonumber \\
\langle 0| J_{J/\psi}^{\mu} (0)|J/\psi(p)\rangle &=&f_{J/\psi}m_{J/\psi}\varepsilon^\mu(p,s) \, ,
\end{eqnarray}
and hadronic coupling constants $G_{P\eta_c\Lambda}$, $G_{P\bar{D}_s\Lambda_c}$,  $G_V$ and $G_T$,
\begin{eqnarray}\label{Coupling-GVGT}
\langle\eta_c(p)\Lambda(q)|P_{cs}(p^\prime)\rangle&=& G_{P\eta_c\Lambda}\,\overline{U}(q)U(p^\prime)\, ,\nonumber\\
\langle \bar{D}_s(p)\Lambda_c(q)|P_{cs}(p^\prime)\rangle&=& G_{P\bar{D}_s\Lambda_c}\,\overline{U}(q)U(p^\prime)\, ,\nonumber\\
\langle J/\psi(p)\Lambda(q)|P_{cs}(p^\prime)\rangle&=& -i\overline{U}(q)\varepsilon^*_\alpha\left( G_V\gamma^\alpha-i\frac{G_T}{m_P+m_\Lambda}\sigma^{\alpha\beta}p_\beta\right)\gamma_5U(p^\prime)\, ,
\end{eqnarray}
 for consistency  with other works, the $U(p^\prime,s)$ and $U(q,s)$ are the Dirac spinors, and the $\varepsilon_\mu$ is the polarization vector.

According to  the quark-hadron duality, we  match the hadron representation with QCD representation of the correlation functions, and
multiply both sides by some Dirac
$\gamma$-matrices $\Gamma^\prime$, then accomplish  the trace in the Dirac spinor space \cite{Duality-Penta-mole-WangZG-CPC-2020},
 \begin{eqnarray}\label{Tr-Gamma}
\frac{1}{4}{\rm Tr}\left[\Pi_{H}(p,q)\Gamma^\prime\right]&=&\frac{1}{4}{\rm Tr}\left[\Pi_{QCD}(p,q)\Gamma^\prime\right]\, ,
\end{eqnarray}
 where   the subscripts $H$ and $QCD$ stand for the  hadron side and   QCD side, respectively.
We set $\Gamma^\prime=\sigma_{\mu\nu}$, $\gamma_5\!\not\!{z}$, $\gamma_5$ in Eq.\eqref{Tr-Gamma}, and carry out  the traces,
\begin{eqnarray}
\frac{1}{4}{\rm Tr}\left[\Pi_{H}(p,q)\sigma_{\mu\nu}\right]&=& \Pi_H(p^{\prime2},p^2,q^2) \,\left(p_\mu q_\nu-q_\mu p_\nu \right)+\cdots\, , \nonumber\\
\frac{1}{4}{\rm Tr}\left[\Pi^\mu_{H}(p,q)\gamma_5\!\not\!{z} \right]&=& \Pi_H^1(p^{\prime2},p^2,q^2)\, q^\mu p\cdot z+\cdots\, ,\nonumber\\
\frac{1}{4}{\rm Tr}\left[\Pi^\mu_{H}(p,q)\gamma_5  \right]&=& \Pi_H^2(p^{\prime2},p^2,q^2)\, q^\mu  +\cdots\, ,
\end{eqnarray}
where the $z_\mu$ is an arbitrary four-vector. We choose the tensor structures $p_\mu q_\nu-q_\mu p_\nu$,  $q_\mu p\cdot z$ and $q_\mu$ to study the hadronic coupling constants,
 \begin{eqnarray}
 \Pi_{H}^{P\eta_c\Lambda}(p^{\prime2},p^2,q^2)   & = & \frac{\lambda_{P\eta_c\Lambda} }{\left(m_P^2-p^{\prime2}\right)\left(m_{\eta_c}^2-p^{2}\right)
 \left(m_{\Lambda}^2-q^{2}\right)} G_{P\eta_c\Lambda}\nonumber\\
 &&+\frac{1 }{\left(m_{\eta_c}^2-p^{2}\right)
 \left(m_{\Lambda}^2-q^{2}\right)} \int_{s_P^0}^\infty ds \frac{\rho_{P^\prime \eta_c\Lambda}(s,p^2,q^2)}{s-p^{\prime2}}\nonumber\\
 &&+\frac{1 }{\left(m_P^2-p^{\prime2}\right)
 \left(m_{\Lambda}^2-q^{2}\right)} \int_{s_{\eta_c^\prime}^0}^\infty ds \frac{\rho_{P \eta_c^\prime  \Lambda}(p^{\prime2},s,q^2)}{s-p^{2}}\nonumber\\
 &&+\frac{1 }{\left(m_P^2-p^{\prime2}\right)\left(m_{\eta_c}^2-p^{2}\right)
 } \int_{s_\Lambda^0}^\infty ds \frac{\rho_{P \eta_c \Lambda^\prime}(p^{\prime2},p^2,s)}{s-q^{2}}+\cdots\, ,
\end{eqnarray}

\begin{eqnarray}
 \Pi_H^{P\bar{D}_s\Lambda_c}(p^{\prime2},p^2,q^2)  & = & \frac{\lambda_{P\bar{D}_s\Lambda_c}}{\left(m_P^2-p^{\prime2}\right)\left(m_{D_s}^2-p^{2}\right)
 \left(m_{\Lambda_c}^2-q^{2}\right)}G_{P\bar{D}_s\Lambda_c}\nonumber\\
 &&+\frac{1 }{\left(m_{D_s}^2-p^{2}\right)
 \left(m_{\Lambda_c}^2-q^{2}\right)} \int_{s_P^0}^\infty ds \frac{\rho_{P^\prime \bar{D}_s\Lambda_c}(s,p^2,q^2)}{s-p^{\prime2}}\nonumber\\
 &&+\frac{1 }{\left(m_P^2-p^{\prime2}\right)
 \left(m_{\Lambda_c}^2-q^{2}\right)} \int_{s_{D_s^\prime}^0}^\infty ds \frac{\rho_{P \bar{D}_s^\prime  \Lambda_c}(p^{\prime2},s,q^2)}{s-p^{2}}\nonumber\\
 &&+\frac{1 }{\left(m_P^2-p^{\prime2}\right)\left(m_{D_s}^2-p^{2}\right)
 } \int_{s_{\Lambda_c}^0}^\infty ds \frac{\rho_{P \bar{D}_s \Lambda_c^\prime}(p^{\prime2},p^2,s)}{s-q^{2}}+\cdots\, ,
\end{eqnarray}

\begin{eqnarray}
 \Pi^{PJ/\psi\Lambda,1}_H(p^{\prime2},p^2,q^2)   & = & \frac{\lambda_{PJ/\psi\Lambda} }{\left(m_P^2-p^{\prime2}\right)\left(m_{J/\psi}^2-p^{2}\right)
 \left(m_{\Lambda}^2-q^{2}\right)} \left(-G_{V}+G_{T}\right)\nonumber\\
 &&+\frac{1 }{\left(m_{J/\psi}^2-p^{2}\right)
 \left(m_{\Lambda}^2-q^{2}\right)} \int_{s_P^0}^\infty ds \frac{\rho^1_{P^\prime J/\psi\Lambda}(s,p^2,q^2)}{s-p^{\prime2}}\nonumber\\
 &&+\frac{1 }{\left(m_P^2-p^{\prime2}\right)
 \left(m_{\Lambda}^2-q^{2}\right)} \int_{s_{\psi^\prime}^0}^\infty ds \frac{\rho^1_{P \psi^\prime  \Lambda}(p^{\prime2},s,q^2)}{s-p^{2}}\nonumber\\
 &&+\frac{1 }{\left(m_P^2-p^{\prime2}\right)\left(m_{J/\psi}^2-p^{2}\right)
 } \int_{s_\Lambda^0}^\infty ds \frac{\rho^1_{P J/\psi \Lambda^\prime}(p^{\prime2},p^2,s)}{s-q^{2}}+\cdots\, ,
\end{eqnarray}

\begin{eqnarray}
\Pi^{PJ/\psi\Lambda,2}_H(p^{\prime2},p^2,q^2)   & = & \frac{\lambda_{PJ/\psi\Lambda}}{\left(m_P^2-p^{\prime2}\right)\left(m_{J/\psi}^2-p^{2}\right)
 \left(m_{\Lambda}^2-q^{2}\right)} \left[\left( m_P-m_\Lambda\right)G_{V}-\frac{m^2_{J/\psi}}{m_P+m_\Lambda}G_{T}\right]\nonumber\\
&&+\frac{1 }{\left(m_{J/\psi}^2-p^{2}\right)
 \left(m_{\Lambda}^2-q^{2}\right)} \int_{s_P^0}^\infty ds \frac{\rho^2_{P^\prime J/\psi\Lambda}(s,p^2,q^2)}{s-p^{\prime2}}\nonumber\\
 &&+\frac{1 }{\left(m_P^2-p^{\prime2}\right)
 \left(m_{\Lambda}^2-q^{2}\right)} \int_{s_{\psi^\prime}^0}^\infty ds \frac{\rho^2_{P \psi^\prime  \Lambda}(p^{\prime2},s,q^2)}{s-p^{2}}\nonumber\\
 &&+\frac{1 }{\left(m_P^2-p^{\prime2}\right)\left(m_{J/\psi}^2-p^{2}\right)
 } \int_{s_\Lambda^0}^\infty ds \frac{\rho^2_{P J/\psi \Lambda^\prime}(p^{\prime2},p^2,s)}{s-q^{2}}+\cdots\, ,
\end{eqnarray}
where $\lambda_{P\eta_c\Lambda}=\frac{f_{\eta_c}m_{\eta_c}^2\lambda_P\lambda_\Lambda}{2m_c}$,
 $\lambda_{P\bar{D}_s\Lambda_c}=\frac{f_{D_s}m_{D_s}^2\lambda_P\lambda_{\Lambda_c}}{m_c}$,
  $\lambda_{PJ/\psi\Lambda}=f_{J/\psi}m_{J/\psi}\lambda_P\lambda_\Lambda$.
The formal  functions $\rho_{P^\prime \eta_c\Lambda}$,
 $\rho_{P \eta_c^\prime  \Lambda}$,  $\rho_{P \eta_c \Lambda^\prime}$,
$\rho_{P^\prime \bar{D}_s\Lambda_c}$,  $\rho_{P \bar{D}_s^\prime  \Lambda_c}$,
 $\rho_{P \bar{D}_s \Lambda_c^\prime}$, $\rho^{1/2}_{P^\prime J/\psi\Lambda}$,
 $\rho^{1/2}_{P \psi^\prime  \Lambda}$,  and
 $\rho^{1/2}_{P J/\psi \Lambda^\prime}$ parameterize the complex transitions among the radial excited states (also P-wave excited states, continuum states) and ground states, their explicit expressions  are difficult to obtain.

We rewrite the correlation functions  $\Pi_H(p^{\prime 2},p^2,q^2)$ as
\begin{eqnarray}
\Pi_{H}(p^{\prime 2},p^2,q^2)&=&\int_{\Delta_{s^\prime}^2}^{s_{P}^0}ds^\prime \int_{\Delta_s^2}^{s^0_{M}}ds \int_{\Delta_u^2}^{s^0_{B}}du  \frac{\rho_H(s^\prime,s,u)}{(s^\prime-p^{\prime2})(s-p^2)(u-q^2)}\nonumber\\
&&+\int_{s^0_P}^{\infty}ds^\prime \int_{\Delta_s^2}^{s^0_{M}}ds \int_{\Delta_u^2}^{s^0_{B}}du  \frac{\rho_H(s^\prime,s,u)}{(s^\prime-p^{\prime2})(s-p^2)(u-q^2)}+\cdots\, ,
\end{eqnarray}
 through triple  dispersion relation, where the hadronic spectral densities $\rho_H(s^\prime,s,u)$,
\begin{eqnarray}\label{Hadron-Spectral-Denti}
\rho_H(s^\prime,s,u)&=&{\lim_{\epsilon_3\to 0}}\,\,{\lim_{\epsilon_2\to 0}} \,\,{\lim_{\epsilon_1\to 0}}\,\,\frac{ {\rm Im}_{s^\prime}\, {\rm Im}_{s}\,{\rm Im}_{u}\,\Pi_H(s^\prime+i\epsilon_3,s+i\epsilon_2,u+i\epsilon_1) }{\pi^3} \, ,
\end{eqnarray}
 the $\Delta_{s^\prime}^2$, $\Delta_s^2$ and $\Delta_u^2$ are the thresholds,  and the  $s_{P}^0$, $s_{M}^0$, $s_{B}^0$ are the continuum thresholds.

On the QCD side, we neglect radiative corrections to the perturbative terms, and  take account of the quark-gluon operators up to dimension $10$ and order $\mathcal{O}( \alpha_s^{k})$ with $k\leq 1$ consistently.
There are two color-neutral clusters for each correlation function, the two clusters translate the four momenta $p_\mu$ (from $0$ to $x$) and $q_\mu$ (from $0$ to $y$) respectively. For the $\eta_c\Lambda$ and $J/\psi\Lambda$ channels, one color-neutral cluster has three light-quark lines, the other has two heavy-quark lines, the three light-quark lines can only contribute two quark-antiquark pairs,   the largest quark-gluon operators are of the form $\bar{q}q\bar{q}q g_sG_{\mu\nu}g_sG_{\alpha\beta}$ with $q=u$, $d$ or $s$, which are of dimension 10. For consistence, we truncate  the operator product expansion up to the vacuum condensates of dimension 10.
In  case of the two-point correlation functions for the hidden-charm pentaquark states, there only exists  one color-neutral cluster connecting $0$ and $x$, the three light-quark lines can contribute three quark-antiquark pairs, while the two heavy quark lines can contribute two gluon-field tensor operators, we obtain  quark-gluon operators of the form $\bar{q}q\bar{q}q \bar{q}qg_sG_{\mu\nu}g_sG_{\alpha\beta}$, which are of dimension 13.
  In addition,  we  take account of all the terms  $\propto m_s$ to account for the light-flavor   $SU(3)$ mass-breaking effects as usual \cite{WangZG-review}. Then we  extract  the vacuum expectations by adopting vacuum saturation routinely.
However,   we cannot write the correlation functions  $\Pi_{QCD}(p^{\prime 2},p^2,q^2)$ in the form,
\begin{eqnarray}
\Pi_{QCD}(p^{\prime 2},p^2,q^2)&=&\int_{\Delta_{s^\prime}^2}^{s_{P}^0}ds^\prime \int_{\Delta_s^2}^{s^0_{M}}ds \int_{\Delta_u^2}^{s^0_{B}}du  \frac{\rho_{QCD}(s^\prime,s,u)}{(s^\prime-p^{\prime2})(s-p^2)(u-q^2)}\nonumber\\
&&+\int_{s^0_P}^{\infty}ds^\prime \int_{\Delta_s^2}^{s^0_{M}}ds \int_{\Delta_u^2}^{s^0_{B}}du  \frac{\rho_{QCD}(s^\prime,s,u)}{(s^\prime-p^{\prime2})(s-p^2)(u-q^2)}+\cdots\, ,
\end{eqnarray}
through triple dispersion relation analogously, as the QCD spectral densities $\rho_{QCD}(s^\prime,s,u)$ cannot exist,
\begin{eqnarray}
\rho_{QCD}(s^\prime,s,u)&=&{\lim_{\epsilon_3\to 0}}\,\,{\lim_{\epsilon_2\to 0}} \,\,{\lim_{\epsilon_1\to 0}}\,\,\frac{ {\rm Im}_{s^\prime}\, {\rm Im}_{s}\,{\rm Im}_{u}\,\Pi_{QCD}(s^\prime+i\epsilon_3,s+i\epsilon_2,u+i\epsilon_1) }{\pi^3} \nonumber\\
&=&0\, ,
\end{eqnarray}
although the correlation functions  $\Pi_{QCD}(p^{\prime 2},p^2,q^2)$ depend on the $p^{\prime 2}$.
We have no choice  but to write the correlation functions  $\Pi_{QCD}(p^{\prime 2},p^2,q^2)$  in the following form,
\begin{eqnarray}
\Pi_{QCD}(p^{\prime 2},p^2,q^2)&=&  \int_{\Delta_s^2}^{s^0_{M}}ds \int_{\Delta_u^2}^{s^0_{B}}du  \frac{\rho_{QCD}(p^{\prime2},s,u)}{(s-p^2)(u-q^2)}+\cdots\, ,
\end{eqnarray}
through double (rather than triple) dispersion relation, where
\begin{eqnarray}
\rho_{QCD}(p^{\prime 2},s,u)&=& {\lim_{\epsilon_2\to 0}} \,\,{\lim_{\epsilon_1\to 0}}\,\,\frac{  {\rm Im}_{s}\,{\rm Im}_{u}\,\Pi_{QCD}(p^{\prime 2},s+i\epsilon_2,u+i\epsilon_1) }{\pi^2} \, .
\end{eqnarray}
Thereafter we would like to write   $\rho_{QCD}(p^{\prime 2},s,u)$ as  $\rho_{QCD}(s,u)$ for simplicity.

We have no choice  but to match the $\Pi_{H}(p^{\prime2},p^2,q^2)$  with  $\Pi_{QCD}(p^{\prime2},p^2,q^2)$ below the
  continuum thresholds $s_M^0$ and $s_B^0$ simultaneously
   to obtain the rigorous  duality \cite{Duality-WangZG-EPJC-2016,
Duality-WangZG-ZhangJX-EPJC-2018,Duality-WangZG-EPJC-2019},
\begin{eqnarray}\label{duality}
\int_{\Delta_s^2}^{s_M^0} ds \int_{\Delta_u^2}^{s_B^0} du \frac{1}{(s-p^2)(u-q^2)}\left[ \int_{\Delta_{s^\prime}^2}^{\infty} ds^\prime \frac{\rho_{H}(s^{\prime},s,u)}{s^\prime-p^{\prime2}}\right]&=&\int_{\Delta_s^2}^{s_M^0} ds \int_{\Delta_u^2}^{s_B^0} du \frac{\rho_{QCD}(s,u)}{(s-p^2)(u-q^2)}\, , \nonumber\\
\end{eqnarray}
we use the word "rigorous duality", because we implement one to one correspondence in the $s$ and $u$ channels, the hadrons in the $s^\prime$ channel have no correspondence on the QCD side and are integrated over.
We accomplish the integral $ ds^\prime $ according to
 the  spectral densities $\rho_{H}(s^{\prime},s,u)$ in Eq.\eqref{Hadron-Spectral-Denti}, then  write down  the relevant expressions explicitly,
\begin{eqnarray}\label{Dual-etaLambda}
  \int_{4m_c^2}^{s^0_{\eta_c}}ds \int_{0}^{s^0_{\Lambda}}du  \frac{\rho^{\eta_c\Lambda}_{QCD}(s,u)}{(s-p^2)(u-q^2)}
  &=&\frac{\lambda_{P\eta_c\Lambda}G_{P\eta_c\Lambda}}{\left(m_P^2-p^{\prime2}\right)\left(m_{\eta_c}^2-p^{2}\right)
  \left(m_{\Lambda}^2-q^{2}\right)}\nonumber\\
    &&+\frac{C_{P^{\prime}\eta_c\Lambda}}{(m_{\eta_c}^2-p^{2})(m_{\Lambda}^2-q^2)} \, ,
\end{eqnarray}

\begin{eqnarray}\label{Dual-DsLambdac}
  \int_{m_c^2}^{s^0_{D_s}}ds \int_{m_c^2}^{s^0_{\Lambda_c}}du  \frac{\rho^{\bar{D}_s\Lambda_c}_{QCD}(s,u)}{(s-p^2)(u-q^2)}
  &=&\frac{\lambda_{P\bar{D}_s\Lambda_c}G_{P\bar{D}_s\Lambda_c}}
  {\left(m_P^2-p^{\prime2}\right)\left(m_{D_s}^2-p^{2}\right)
  \left(m_{\Lambda_c}^2-q^{2}\right)}\nonumber\\
    &&+\frac{C_{P^{\prime}\bar{D}_s\Lambda_c}}{(m_{D_s}^2-p^{2})(m_{\Lambda_c}^2-q^2)} \, ,
\end{eqnarray}

\begin{eqnarray}\label{Dual-JpsiLambda-1}
  \int_{4m_c^2}^{s^0_{J/\psi}}ds \int_{0}^{s^0_{\Lambda}}du  \frac{\rho^{J/\psi \Lambda,1}_{QCD}(s,u)}{(s-p^2)(u-q^2)}
  &=&\lambda_{PJ/\psi\Lambda} \frac{G_T-G_V}{\left(m_P^2-p^{\prime2}\right)
  \left(m_{J/\psi}^2-p^{2}\right)\left(m_{\Lambda}^2-q^{2}\right)}\nonumber\\
    &&+\frac{C_{P^{\prime}J/\psi\Lambda,1}}{(m_{J/\psi}^2-p^{2})(m_{\Lambda}^2-q^2)} \, ,
\end{eqnarray}

 \begin{eqnarray}\label{Dual-JpsiLambda-2}
  \int_{4m_c^2}^{s^0_{J/\psi}}ds \int_{0}^{s^0_{\Lambda}}du  \frac{\rho^{J/\psi \Lambda,2}_{QCD}(s,u)}{(s-p^2)(u-q^2)}
  &=&\lambda_{PJ/\psi\Lambda} \frac{\left(m_P-m_\Lambda\right)G_V-G_T\frac{m_{J/\psi}^2}{m_P+
  m_\Lambda}}{\left(m_P^2-p^{\prime2}\right)\left(m_{J/\psi}^2-p^{2}\right)
  \left(m_{\Lambda}^2-q^{2}\right)}\nonumber\\
  &&+\frac{C_{P^{\prime}J/\psi\Lambda,2}}{(m_{J/\psi}^2-p^{2})(m_{\Lambda}^2-q^2)} \, ,
\end{eqnarray}
where
\begin{eqnarray}\label{define-C-etacLambda}
 C_{P^\prime\eta_c\Lambda}   & = &  \int_{s_P^0}^\infty ds \frac{\rho_{P^\prime \eta_c\Lambda}(s,p^2,q^2)}{s-p^{\prime2}} \, ,
\end{eqnarray}

\begin{eqnarray}\label{define-C-DsLambdac}
 C_{P^\prime \bar{D}_s\Lambda_c}  & = & \int_{s_P^0}^\infty ds \frac{\rho_{P^\prime \bar{D}_s\Lambda_c}(s,p^2,q^2)}{s-p^{\prime2}}\, ,
\end{eqnarray}

\begin{eqnarray}\label{define-C-JpsiLambda}
 C_{P^\prime J/\psi\Lambda,1/2}
 &=&\int_{s_P^0}^\infty ds \frac{\rho_{P^\prime J/\psi\Lambda}^{1/2}(s,p^2,q^2)}{s-p^{\prime2}}\, .
\end{eqnarray}
We take the parameters $C_{P^{\prime}\eta_c\Lambda}$, $C_{P^{\prime}\bar{D}_s\Lambda_c}$ and $C_{P^{\prime}J/\psi\Lambda,1/2}$ as free parameters and search for the optimal values to obtain stable QCD sum rules.

The current $J_P(x)$ couples potentially  both to the pentaquark candidate $P_{cs}(4338)$ and meson-baryon pairs $J/\psi \Lambda$, $\eta_c\Lambda$, $\bar{D}_s\Lambda_c$, etc, if they have the same quantum numbers.  In the two-point QCD sum rules, there are both pentaquark and meson-baryon pair contributions, the single-pole term (or pentaquark state) dominates the QCD sum rules, the  meson-baryon scattering states only contribute a finite width to modify the dispersion relation \cite{WangZG-penta-Jpsi-Lamba,WangZG-Landau-PRD-2020}. In the present case, the three-point QCD sum rules, on the QCD side,  there exist both connected and disconnected Feynman diagrams, the two color-neutral clusters, which connect
$0$ to $x$ and $0$ to $y$ respectively in the coordinate space, amplify the factorizable  contributions. The dominant contributions come from the disconnected Feynman diagrams and make contributions both to the pentaquark states and baryon-meson pairs, the minor contributions come from the connected Feynman diagrams, which only contribute  to the pentaquark states. In the present work,
we would  take account of the meson-baryon pair contributions,
\begin{eqnarray}
\langle 0|J_P(0)|\eta_c(p)\Lambda(q)\rangle&=& \lambda_{\eta_c\Lambda}\,U(q)\, , \nonumber\\
\langle 0|J_P(0)|\bar{D}_s(p)\Lambda_c(q)\rangle&=& \lambda_{\bar{D}_s\Lambda_c}U(q)\, , \nonumber\\
\langle 0|J_P(0)|J/\psi(p)\Lambda(q)\rangle&=& \lambda_{J/\psi\Lambda}^V\,\!\not\!{\varepsilon}\gamma_5\, U(q)+\lambda_{J/\psi\Lambda }^T\,\varepsilon_\alpha \sigma^{\alpha\beta}p_\beta\gamma_5\, U(q)\, ,
\end{eqnarray}
where the $\lambda_{\eta_c\Lambda}$, $\lambda_{\bar{D}_s\Lambda_c}$, $\lambda_{J/\psi\Lambda}^V$ and $\lambda_{J/\psi\Lambda }^T$ are the coupling constants, and lead to additional contributions having the pole structures,
\begin{eqnarray}\label{C-BM-pair}
 \frac{1}{(m_{\eta_c}^2-p^{2})(m_{\Lambda}^2-q^2)} \,  , \,\, \, \frac{1}{(m_{\bar{D}_s}^2-p^{2})(m_{\Lambda_c}^2-q^2)} \, , \,\, \,\frac{1}{(m_{J/\psi}^2-p^{2})(m_{\Lambda}^2-q^2)} \, ,
\end{eqnarray}
in Eqs.\eqref{Dual-etaLambda}-\eqref{Dual-JpsiLambda-2} sequentially.
The contributions could be absorbed into the coefficients $C_{P^{\prime}\eta_c\Lambda}$, $C_{P^{\prime}\bar{D}_s\Lambda_c}$ and $C_{P^{\prime}J/\psi\Lambda,1/2}$, irrespective of their magnitudes \cite{WangZG-review}.

We set $p^{\prime2}=p^2$ for the  $P \eta_c\Lambda$ and $P J/\psi\Lambda$ channels and $p^{\prime2}=\alpha q^2$  for the  $P \bar{D}_s\Lambda_c$ channel.
 Then we perform  double Borel transformation in regard  to the variables $P^2=-p^2$ and $Q^2=-q^2$ respectively, and set $T_1^2=T_2^2=T^2$ to obtain four   QCD sum rules,
\begin{eqnarray}\label{QCDSR-etacN}
&& \frac{\lambda_{P\eta_c\Lambda}G_{P\eta_c\Lambda}}{m_{P}^2-m_{\eta_c}^2} \left[ \exp\left(-\frac{m_{\eta_c}^2}{T^2} \right)-\exp\left(-\frac{m_{P}^2}{T^2} \right)\right]\exp\left(-\frac{m_{\Lambda}^2}{T^2} \right) +\nonumber\\
&&C_{P^{\prime}\eta_c\Lambda} \exp\left(-\frac{m_{\eta_c}^2+m_{\Lambda}^2}{T^2} \right)=\int_{4m_c^2}^{s_{\eta_c}^0} ds \int_{0}^{s_\Lambda^0} du\, \rho^{\eta_c\Lambda}_{QCD}(s,u)\exp\left(-\frac{s+u}{T^2}  \right)\, ,
\end{eqnarray}

\begin{eqnarray}\label{QCDSR-DsLambdac}
&& \frac{\lambda_{P\bar{D}_s\Lambda_c}G_{P\bar{D}_s\Lambda_c}}
{\alpha\left(\widetilde{m}_{P}^2-m_{\Lambda_c}^2\right)} \left[ \exp\left(-\frac{m_{\Lambda_c}^2}{T^2} \right)-\exp\left(-\frac{\widetilde{m}_{P}^2}{T^2} \right)\right]\exp\left(-\frac{m_{D_s}^2}{T^2} \right) +\nonumber\\
&&C_{P^{\prime}\bar{D}_s\Lambda_c} \exp\left(-\frac{m_{\Lambda_c}^2+m_{D_s}^2}{T^2} \right)=\int_{m_c^2}^{s_{D_s}^0} ds \int_{m_c^2}^{s_{\Lambda_c}^0} du\, \rho^{\bar{D}_s\Lambda_c}_{QCD}(s,u)\exp\left(-\frac{s+u}{T^2}  \right)\, ,
\end{eqnarray}

\begin{eqnarray}\label{QCDSR-GV-GT}
&& \frac{\lambda_{PJ/\psi\Lambda}G_{V/T}}{m_{P}^2-m_{J/\psi}^2} \left[ \exp\left(-\frac{m_{J/\psi}^2}{T^2} \right)-\exp\left(-\frac{m_{P}^2}{T^2} \right)\right]\exp\left(-\frac{m_{\Lambda}^2}{T^2} \right) +\nonumber\\
&&C_{V/T} \exp\left(-\frac{m_{J/\psi}^2+m_{\Lambda}^2}{T^2} \right)=\int_{4m_c^2}^{s_{J/\psi}^0} ds \int_{0}^{s_\Lambda^0} du\, \rho^{V/T}_{QCD}(s,u)\exp\left(-\frac{s+u}{T^2}  \right)\, ,
\end{eqnarray}

\begin{eqnarray}
C_V&=&\left[\frac{m_{J/\psi}^2}{m_P+m_\Lambda}C_{P^{\prime} J/\psi\Lambda,1} + C_{P^{\prime} J/\psi\Lambda,2}  \right]\frac{m_P+m_\Lambda}{m_P^2-m_{J/\psi}^2-m_\Lambda^2}\, ,\nonumber\\
C_T&=&\left[\left(m_P-m_\Lambda\right)C_{P^{\prime} J/\psi\Lambda,1}+C_{P^{\prime} J/\psi\Lambda,2} \right]\frac{m_P+m_\Lambda}{m_P^2-m_{J/\psi}^2-m_\Lambda^2}\, ,\nonumber\\
\rho_{QCD}^V(s,u)&=&\left[\frac{m_{J/\psi}^2}{m_P+m_\Lambda}\rho_{QCD}^{J/\psi \Lambda,1}(s,u)+\rho_{QCD}^{J/\psi \Lambda,2}(s,u) \right]\frac{m_P+m_\Lambda}{m_P^2-m_{J/\psi}^2-m_\Lambda^2}\, ,\nonumber\\
\rho_{QCD}^T(s,u)&=&\left[\left(m_P-m_\Lambda\right)\rho_{QCD}^{J/\psi \Lambda,1}(s,u)+\rho_{QCD}^{J/\psi \Lambda,2}(s,u) \right]\frac{m_P+m_\Lambda}{m_P^2-m_{J/\psi}^2-m_\Lambda^2}\, ,
\end{eqnarray}
$\tilde{m}_P^2=m_P^2/\alpha$, and adopt the value $\alpha=2$ as in Refs.\cite{Duality-WangZG-AAPPS-2024,Duality-YangXS-CPC-2025}, the lengthy  expressions of the spectral densities $\rho^{\eta_c\Lambda}_{QCD}(s,u)$, $\rho^{\bar{D}_s\Lambda_c}_{QCD}(s,u)$ and $\rho^{J/\psi \Lambda,1/2}_{QCD}(s,u)$ are given clearly in the Appendix.  In Ref.\cite{Duality-WangXW-CPC-2024}, we took the $T_1^2$ and $T_2^2$ as two independent parameters, and explored   the 3D surfaces of the hadronic coupling constants $G=G(T_1^2,T_2^2)$ for the first time. We determined the Borel platforms by setting $T^2_{1, max}-T^2_{1,min}=1\,{\rm GeV^2}$ and $T^2_{2,max}-T^2_{2,min}=1\,{\rm GeV^2}$, where the subscripts
$max$ and $min$ denote the maximum and minimum values, respectively. Then we quantified the "flat surface" via minimizing the average relative error bounds  through optimizing the unknown parameters $C$, and plotted the 3D  graphs. We observed that it is reasonable to set $T_1^2=T_2^2=T^2$ for simplicity in studying the hidden-charm pentaquark states.

\begin{figure}
\centering
\includegraphics[totalheight=8cm,width=10cm]{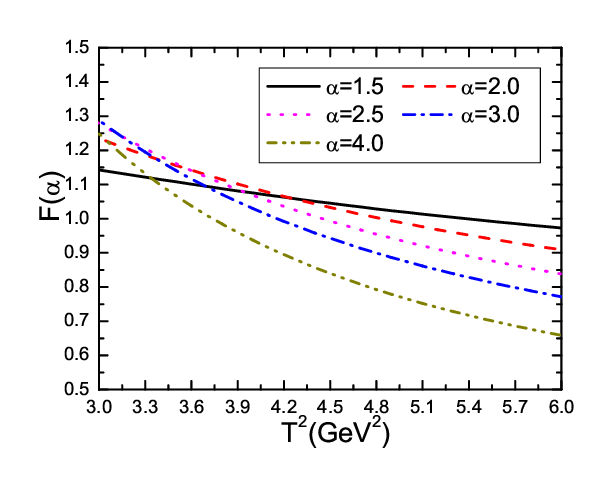}
  \caption{ The factor $F(\alpha)$    with variations of the  Borel parameter $T^2$. }\label{f-afa}
\end{figure}

We can define a function $F(\alpha)$ to study  dependence of the predicted hadronic coupling constant on the parameter $\alpha$,
\begin{eqnarray}
F(\alpha)&=& \frac{1}
{\alpha\left(\widetilde{m}_{P}^2-m_{\Lambda_c}^2\right)} \left[ \exp\left(-\frac{m_{\Lambda_c}^2}{T^2} \right)-\exp\left(-\frac{\widetilde{m}_{P}^2}{T^2} \right)\right]\exp\left(-\frac{m_{D_s}^2}{T^2} \right) \, ,
\end{eqnarray}
as we usually set $p^{\prime2}=\alpha q^2$ before performing the Borel transformation \cite{WangZG-review,Duality-WangZG-ZhangJX-EPJC-2018},   where $\widetilde{m}_{P}^2=m_P^2/\alpha$.

In Fig.\ref{f-afa}, we plot the factor $F(\alpha)$    with variations of the  Borel parameter $T^2$, which is normalized by setting $F(1)=1$. From the figure, we can see clearly that at the region $T^2>4.5\,\rm{GeV}^2$, the $F(\alpha)$ decreases monotonously with increase of the values of the $\alpha$. As  the hadronic  coupling constant $G_{P\bar{D}_s\Lambda_c}\propto 1/F(\alpha)$ from the QCD sum rules in Eq.\eqref{QCDSR-DsLambdac}, a smaller $F(\alpha)$ means a larger $G_{P\bar{D}_s\Lambda_c}$, the masses of the $\bar{D}_s$ and $\Lambda_c$ are about half of the mass of the $P_{cs}(4338)$, just like in the case of the fully-charm tetraquark states, where the masses of the charmonia are about half of the masses of the fully-charm tetraquark states. In Refs.\cite{Duality-WangZG-AAPPS-2024,Duality-YangXS-CPC-2025}, we set $\alpha=2$ to study the hadronic coupling constants of the fully-charm tetraquark states and obtained  suitable decay widths to reproduce the experimental data, other values, such as $\alpha=1$ or $3$, would over-estimate or under-estimate the experimental data  remarkably.
In the present work, the value $\alpha=2$ works and we abandon fine-tuning. Moreover, the relation  $m_{J/\psi}^2/m_{P}^2\approx m_{\Lambda_c}^2/\tilde{m}_{P}^2$ holds approximately, and we warrant to adopt a uniform criterion for consistence. If we choose the normalization $F(\alpha)|_{\alpha=2}=1$ at the centroid of the Borel window, the uncertainties $\delta \alpha=\mp 1$ lead to the uncertainties $\delta F(\alpha)|_{\alpha=2}={}^{+0.03}_{-0.12}$.

In the large $T^2$ limit, the four QCD sum rules are reduced to
\begin{eqnarray}
 \frac{\lambda_{P\eta_c\Lambda}G_{P\eta_c\Lambda}}{T^2} +
C_{P^{\prime}\eta_c\Lambda} &=& {\rm constant}\, , \nonumber\\
 \frac{\lambda_{P\bar{D}_s\Lambda_c}G_{P\bar{D}_s\Lambda_c}}
{2T^2} +C_{P^{\prime}\bar{D}_s\Lambda_c} &=&{\rm constant}\, ,\nonumber\\
 \frac{\lambda_{PJ/\psi\Lambda}G_{V/T}}{T^2}  +
C_{V/T} &=&{\rm constant}\, ,
\end{eqnarray}
or more precisely,
\begin{eqnarray}
 C_{P^{\prime}\eta_c\Lambda} &=& {\rm constant}\, , \nonumber\\
 C_{P^{\prime}\bar{D}_s\Lambda_c} &=&{\rm constant}\, ,\nonumber\\
C_{V/T} &=&{\rm constant}\, ,
\end{eqnarray}
which are dominated by the contributions from the higher resonances and continuum states. We obtain the conclusion that the free parameters $C$ are necessary.  For small and moderate values of the Borel parameter $T^2$, it is reasonable to set the unknown parameters as $C=A+B*T^2$, as $C\to A$ in the limit of small $T^2$.

\section{Numerical results and discussions}
On the hadron side, we take
$m_{J/\psi}=3.0969\,\rm{GeV}$, $m_{\Lambda}=1.115683\,\rm{GeV}$,
$m_{\eta_c}=2.98409\,\rm{GeV}$,  $m_{D_s}=1.969\,\rm{GeV}$, $m_{\Lambda_c}=2.28646\,\rm{GeV}$, $\sqrt{s^0_{D_s}}=2.55\,\rm{GeV}$, $\sqrt{s^0_{J/\psi}}=3.60\,\rm{GeV}$, $\sqrt{s^0_{\eta_c}}=3.50\,\rm{GeV}$    from the Particle Data Group \cite{PDG},
$f_{J/\psi}=0.418 \,\rm{GeV}$, $f_{\eta_c}=0.387 \,\rm{GeV}$  from the coupled lattice QCD and QCD sum rule analysis \cite{Becirevic},
$\sqrt{s^0_{\Lambda}}=1.59\,\rm{GeV}$ \cite{Pcs4338-mole-decay-WangXW-PRD},
$f_{D_s}=0.240 \,\rm{GeV}$  \cite{WangZG-decay-cnstant-D},
$\lambda_\Lambda=0.0287\,\rm{GeV}^3\frac{\sqrt{3}}{2}$ \cite{Pcs4338-mole-decay-WangXW-PRD}, $\sqrt{s^0_{\Lambda_c}}=2.75\,\rm{GeV}$,  $\lambda_{\Lambda_c}=0.022\,\rm{GeV}^3$ \cite{WangZG-Lambdac,WangZG-Lambdac-2S},   $m_{P}=4.33\,\rm{GeV}$, $\lambda_P=2.34\times 10^{-3}\,\rm{GeV}^6$  \cite{WangZG-penta-Jpsi-Lamba} from the two-point QCD sum rules.

On the QCD side, we choose  the  vacuum condensates
$\langle\bar{q}q \rangle=-(0.24\pm 0.01\, \rm{GeV})^3$,  $\langle\bar{s}s \rangle=(0.8\pm0.1)\langle\bar{q}q \rangle$,
 $\langle\bar{q}g_s\sigma G q \rangle=m_0^2\langle \bar{q}q \rangle$, $\langle\bar{s}g_s\sigma G s \rangle=m_0^2\langle \bar{s}s \rangle$,
$m_0^2=(0.8 \pm 0.1)\,\rm{GeV}^2$, $\langle \frac{\alpha_s
GG}{\pi}\rangle=0.012\pm0.004\,\rm{GeV}^4$    at the energy scale  $\mu=1\, \rm{GeV}$  \cite{SVZ79,PRT85,ColangeloReview}, and choose the $\overline{MS}$ quark  masses $m_{c}(m_c)=(1.275\pm0.025)\,\rm{GeV}$
 and $m_s(\mu=2\,\rm{GeV})=(0.095\pm0.005)\,\rm{GeV}$
  from the Particle Data Group \cite{PDG}.
In addition, we consider  the energy-scale dependence of  the input   parameters,
\begin{eqnarray}
 \langle\bar{q}q \rangle(\mu)&=&\langle\bar{q}q\rangle({\rm 1 GeV})\left[\frac{\alpha_{s}({\rm 1 GeV})}{\alpha_{s}(\mu)}\right]^{\frac{12}{33-2n_f}}\, , \nonumber\\
 \langle\bar{s}s \rangle(\mu)&=&\langle\bar{s}s \rangle({\rm 1 GeV})\left[\frac{\alpha_{s}({\rm 1 GeV})}{\alpha_{s}(\mu)}\right]^{\frac{12}{33-2n_f}}\, , \nonumber\\
 \langle\bar{q}g_s \sigma Gq \rangle(\mu)&=&\langle\bar{q}g_s \sigma Gq \rangle({\rm 1 GeV})\left[\frac{\alpha_{s}({\rm 1 GeV})}{\alpha_{s}(\mu)}\right]^{\frac{2}{33-2n_f}}\, ,\nonumber\\
  \langle\bar{s}g_s \sigma Gs \rangle(\mu)&=&\langle\bar{s}g_s \sigma Gs \rangle({\rm 1 GeV})\left[\frac{\alpha_{s}({\rm 1 GeV})}{\alpha_{s}(\mu)}\right]^{\frac{2}{33-2n_f}}\, ,\nonumber\\
m_c(\mu)&=&m_c(m_c)\left[\frac{\alpha_{s}(\mu)}{\alpha_{s}(m_c)}\right]^{\frac{12}{33-2n_f}} \, ,\nonumber\\
m_s(\mu)&=&m_s({\rm 2GeV} )\left[\frac{\alpha_{s}(\mu)}{\alpha_{s}({\rm 2GeV})}\right]^{\frac{12}{33-2n_f}}\, ,\nonumber\\
\alpha_s(\mu)&=&\frac{1}{b_0t}\left[1-\frac{b_1}{b_0^2}\frac{\log t}{t} +\frac{b_1^2(\log^2{t}-\log{t}-1)+b_0b_2}{b_0^4t^2}\right]\, ,
\end{eqnarray}
  where $t=\log \frac{\mu^2}{\Lambda^2}$, $b_0=\frac{33-2n_f}{12\pi}$, $b_1=\frac{153-19n_f}{24\pi^2}$, $b_2=\frac{2857-\frac{5033}{9}n_f+\frac{325}{27}n_f^2}{128\pi^3}$,  $\Lambda_{QCD}=210\,\rm{MeV}$, $292\,\rm{MeV}$  and  $332\,\rm{MeV}$ for the flavors  $n_f=5$, $4$ and $3$, respectively  \cite{PDG,Narison-mix}, and evolve them to an acceptable energy scale $\mu$  with $n_f=4$ to extract the hadronic coupling constants.

  The optimal energy scale of the two-point QCD sum rules for the $P_{cs}(4338)$ as a hidden-charm pentaquark state with the spin-parity $J^P={\frac{1}{2}}^-$ is $\mu=2.2\,\rm{GeV}$ \cite{WangZG-penta-Jpsi-Lamba}, which is determined   by the modified energy scale formula. The energy scale $\mu=2.2\,\rm{GeV}$ is too large  for the mesons $\eta_c$, $\bar{D}_s$, $J/\psi$ and baryons $\Lambda_c$, $\Lambda$.
     And we would like to choose the typical energy scale  $\mu=1.0\,\rm{GeV}$, which is widely accepted for the ground states hadrons.
 At the energy scale $\mu=1.0\,\rm{GeV}$, the central value $m_{c}(1\,{\rm GeV})=1.3855\,\rm{GeV}$, the integral intervals  $\sqrt{s^0_{\eta_c}}-2m_{c}(1\,{\rm GeV})=0.729\,\rm{GeV}< 1\, \rm{GeV}$ and $\sqrt{s^0_{J/\psi}}-2m_{c}(1\,{\rm GeV})=0.829\,\rm{GeV}< 1\, \rm{GeV}$,  the integrals on the QCD side are very sensitive  to the uncertainty $\delta m_c$, which leads to a uncertainty as large as $\pm 0.054\,\rm{GeV}$ for the integral intervals,  we would like to choose the central value only.

We should bear in mind that the pole residue $\lambda_P$ is extracted at the energy scale $\mu=2.2\,\rm{GeV}$, and we should evolve it to the typical energy scale $\mu=1.0\,\rm{GeV}$. As the current $J_P(x)$ couples potentially  to the $P_{cs}(4338)$ and $J/\psi \Lambda$ pair according to the Fierz transformation,
 \begin{eqnarray}
 J_P&=&\frac{\varepsilon^{ijk}}{4\sqrt{2}} \left[ u^T_i C\gamma_\alpha s_j\, d_{k}-d^T_i C\gamma_\alpha s_j\, u_{k}\right]\bar{c}\gamma^\alpha c \nonumber\\
 &&-\frac{i\varepsilon^{ijk}}{4\sqrt{2}} \left[ u^T_i C\gamma_\alpha s_j\, \sigma^{\mu\alpha}d_{k}-d^T_i C\gamma_\alpha s_j\, \sigma^{\mu\alpha}u_{k}\right]\bar{c}\gamma_\mu c+\cdots\, ,
 \end{eqnarray}
 we would like to estimate the anomalous dimension $\gamma_{J_P}\approx \gamma_{J_{J/\psi}}+\gamma_{J_{\Lambda}}=-\frac{2}{9}$ \cite{ZhaoZX-gamma}, and obtain a factor,
\begin{eqnarray}\label{Factor-Gamma}
\left(\frac{\log\frac{1.0\,{\rm GeV}}{\Lambda_{QCD}}}{\log\frac{2.2\,{\rm GeV}}{\Lambda_{QCD}}}\right)^{-\frac{2}{9}}&=&1.116\, ,
\end{eqnarray}
to enhance the $\lambda_P$, therefore the hadronic coupling constants $G$ obtain a suppression factor $1/1.116$.

If we choose the value  of the decay constant $f_{D_s}$ from the lattice QCD, $f_{D_s}=249.9\,\rm{MeV}$ \cite{Latt-Decay-fDs}, and take account of the small quark mass $m_s$ via the replacement $f_{D_s}m_{D_s}^2/m_c\to f_{D_s}m_{D_s}^2/(m_c+m_s)$ in defining the decay constant, the
$G_{P\bar{D}_s\Lambda_c}$ undergoes  an enhancement  by a factor,
\begin{eqnarray}\label{Factor-ms-decay}
\frac{0.240/1.3855}{0.2499/(1.3855+0.1161)}&=&1.041\, ,
\end{eqnarray}
as we choose the energy scale $\mu=1\,\rm{GeV}$.

In Ref.\cite{Pcs4338-mole-decay-WangXW-PRD}, we chose the current,
\begin{eqnarray}
\tilde{J}_{\Lambda}(x)&=&  \sqrt{\frac{2}{3}}\varepsilon^{ijk} \left[ u^T_i(x) C\gamma_\alpha s_j(x)\, \gamma^\alpha\gamma_5 d_{k}(x)-d^T_i(x) C\gamma_\alpha s_j(x)\, \gamma^\alpha\gamma_5 u_{k}(x)\right] \, ,
\end{eqnarray}
to  study the mass and pole residue of the $\Lambda$ baryon with the two-point QCD  sum rules,   where $J_{\Lambda}(x)=\frac{\sqrt{3}}{2}\tilde{J}_{\Lambda}(x)$. The current $\tilde{J}_{\Lambda}(x)$ couples potentially  to  the baryons with both positive and negative parities, as multiplying $i\gamma_5$ changes the parity of the current. However, for the Ioffe-type currents interpolating the light baryons, we cannot separate the contributions of the baryons with positive and negative parities unambiguously.   We obtained two QCD sum rules with respect to the Dirac structures $\!\not\!{p}$ and $1$, respectively, and took  the averages  of the predicted mass (reproduces the experimental value) and pole residue \cite{Pcs4338-mole-decay-WangXW-PRD}. If we take the $\Lambda(1405)$  with the spin-parity $J^P={\frac{1}{2}}^-$ as the 1P excitation of the $\Lambda$, we obtain the mass gap $m_{\Lambda(1405)}-m_{\Lambda}=289\, \rm{MeV}$ \cite{PDG}. If we choose the continuum threshold parameter $\sqrt{s^0_{\Lambda}}=1.59\,\rm{GeV}$ \cite{Pcs4338-mole-decay-WangXW-PRD}, contamination from the $\Lambda(1405)$ is included in. However, if we choose a smaller $\sqrt{s^0_{\Lambda}}$, we cannot reproduce the experimental mass. On the other hand, if we choose an enough large $\sqrt{s^0_{\Lambda}}$ to include all the contribution of the 1P state to perform coupled channels analysis, contamination from  the 2S state is included in. Step by step, we should include all the S-wave and P-wave $\Lambda$ baryons  on the hadron side. It is beyond our ability. The mass gap between the $\Lambda_c(2595)$ with the  $J^P={\frac{1}{2}}^-$ and $\Lambda_c$ with the $J^P={\frac{1}{2}}^+$ is also small, $m_{\Lambda_c(2595)}-m_{\Lambda_c}=306\, \rm{MeV}$ \cite{PDG}, analogous ansatz holds for the three-point QCD sum rules, although in the two-point QCD sum rules, we can separate the contributions of the heavy baryons with positive and negative parities  unambiguously for the Ioffe-currents.

If we insist on including the contributions of the $\Lambda(1405)$ and $\Lambda_c(2595)$ as the 1P states, the hadron sides  in Eqs.\eqref{Dual-etaLambda}-\eqref{Dual-JpsiLambda-2} receive additional terms,
\begin{eqnarray}
\frac{\lambda_{P\eta_c\tilde{\Lambda}}G_{P\eta_c\tilde{\Lambda}}}{\left(m_P^2-p^{\prime2}\right)\left(m_{\eta_c}^2-p^{2}\right)
  \left(m_{\tilde{\Lambda}}^2-q^{2}\right)}\, , \,\,\,
\frac{\lambda_{P\bar{D}_s\tilde{\Lambda}_c}G_{P\bar{D}_s\tilde{\Lambda}_c}}
  {\left(m_P^2-p^{\prime2}\right)\left(m_{D_s}^2-p^{2}\right)
  \left(m_{\tilde{\Lambda}_c}^2-q^{2}\right)}\, ,
\end{eqnarray}
\begin{eqnarray}
\lambda_{PJ/\psi\tilde{\Lambda}} \frac{-\tilde{G}_V-\tilde{G}_T\frac{m_P-
  m_{\tilde{\Lambda}}}{m_P+
  m_{\tilde{\Lambda}}}}{\left(m_P^2-p^{\prime2}\right)
  \left(m_{J/\psi}^2-p^{2}\right)\left(m_{\tilde{\Lambda}}^2-q^{2}\right)}\, ,\,\,\,
\lambda_{PJ/\psi\tilde{\Lambda}} \frac{\left(m_P+m_{\tilde{\Lambda}}\right)\tilde{G}_V+\tilde{G}_T
\frac{m_{J/\psi}^2}{m_P+
  m_{\tilde{\Lambda}}}}{\left(m_P^2-p^{\prime2}\right)\left(m_{J/\psi}^2-p^{2}\right)
  \left(m_{\tilde{\Lambda}}^2-q^{2}\right)}\, ,
\end{eqnarray}
which lead to additional terms of the forms,
\begin{eqnarray}
 \left[ \exp\left(-\frac{m_{\eta_c}^2}{T^2} \right)-\exp\left(-\frac{m_{P}^2}{T^2} \right)\right]\exp\left(-\frac{m_{\tilde{\Lambda}}^2}{T^2} \right) \, ,\,\,\,
  \left[ \exp\left(-\frac{m_{\tilde{\Lambda}_c}^2}{T^2} \right)-\exp\left(-\frac{\widetilde{m}_{P}^2}{T^2} \right)\right]\exp\left(-\frac{m_{D_s}^2}{T^2} \right) \, ,\nonumber
\end{eqnarray}
\begin{eqnarray}
  \left[ \exp\left(-\frac{m_{J/\psi}^2}{T^2} \right)-\exp\left(-\frac{m_{P}^2}{T^2} \right)\right]\exp\left(-\frac{m_{\tilde{\Lambda}}^2}{T^2} \right) \, ,
\end{eqnarray}
(after performing Borel transformations) in the four QCD sum rules, see Eqs.\eqref{QCDSR-etacN}-\eqref{QCDSR-GV-GT}, where we represent the $\Lambda(1405)$ and $\Lambda_c(2595)$ as the $\tilde{\Lambda}$ and $\tilde{\Lambda}_c$ respectively for simplicity, and other symbols   are defined analogously according to Eqs.\eqref{QCDSR-etacN}-\eqref{QCDSR-GV-GT}.
We can rewrite them in the following forms,
\begin{eqnarray}\label{1P-etacLambda}
 \left[ 1 -\exp\left(-\frac{m_{P}^2-m_{\eta_c}^2}{T^2} \right)\right]\exp\left(-\frac{m_{\tilde{\Lambda}}^2-m_{\Lambda}^2}{T^2} \right)\exp\left(-\frac{m_{\eta_c}^2+m_{\Lambda}^2}{T^2}\right)
  \, ,
\end{eqnarray}
\begin{eqnarray}\label{1P-DsLambdac}
  \left[ \exp\left(-\frac{m_{\tilde{\Lambda}_c}^2-m_{\Lambda_c}^2}{T^2} \right)-\exp\left(-\frac{\widetilde{m}_{P}^2-m_{\Lambda_c}^2}{T^2} \right)\right]\exp\left(-\frac{m_{D_s}^2+m_{\Lambda_c}^2}{T^2} \right) \, ,
\end{eqnarray}
\begin{eqnarray}\label{1P-JpsiLambda}
 \left[ 1 -\exp\left(-\frac{m_{P}^2-m_{J/\psi}^2}{T^2} \right)\right]\exp\left(-\frac{m_{\tilde{\Lambda}}^2-m_{\Lambda}^2}{T^2} \right)\exp\left(-\frac{m_{J/\psi}^2+m_{\Lambda}^2}{T^2}\right)
  \, ,
\end{eqnarray}
and absorb their contributions into the coefficients $C_{P^{\prime}\eta_c\Lambda}$,
$ C_{P^{\prime}\bar{D}_s\Lambda_c}$ and  $C_{V/T}$ with a redefinition, the contaminations would not harm the predictions. Direct calculations indicate that such an ansatz is not bad. In the three-point QCD sum rules, as the reckless cut of the continuum thresholds $s_0$ of the 1S states sets  in, the $\lambda_{\tilde{\Lambda}}$ and $\lambda_{\tilde{\Lambda}_c}$  cannot appear as their physical values via receiving enough contributions,  the values of the pole residues $\lambda_{\tilde{\Lambda}}$ and $\lambda_{\tilde{\Lambda}_c}$ are expected to be small.

We would like to  choose $\xi$ to stand for the input parameters  on the QCD side,  in the two-point QCD sum rules, the uncertainties $\delta\xi$ lead to  uncertainties of the hadron masses and pole residues (decay constants), we could separate the uncertainties $\delta m_{P}$, $\delta m_{M}$, $\delta m_{B}$, $\delta \lambda_{P}$, $\delta \lambda_{M}$ and $\delta \lambda_{B}$ unambiguously. In the three-point QCD sum rules,
the uncertainties   $\delta \xi$ give rise to the uncertainties $\delta\,\lambda_{P}\lambda_{M}\lambda_{B}G_{PMB}$ and $\delta C_{P^{\prime}MB}$,
where
\begin{eqnarray}\label{Uncertainty-4}
\delta\,\lambda_{P}\lambda_{M}\lambda_{B}G_{PMB} &=&\lambda_{P}\lambda_{M}\lambda_{B}G_{PMB}\left( \frac{\delta \lambda_{P}}{\lambda_{P}} +\frac{\delta \lambda_{M}}{\lambda_{M}}+\frac{\delta \lambda_{B}}{\lambda_{B}}+\frac{\delta G_{PMB}}{G_{PMB}}\right)\, ,
\end{eqnarray}
and $\lambda_{M}=\frac{f_{\eta_c}m_{\eta_c}^2}{2m_c}$,
$\frac{f_{D_s}m_{D_s}^2}{m_c}$, $f_{J/\psi}m_{J/\psi}$. The uncertainties
$\delta\,\lambda_{P}\lambda_{M}\lambda_{B}G_{PMB}$ appear as collective effects, and we could not separate the components $\frac{\delta \lambda_{P}}{\lambda_{P}}$, $\frac{\delta \lambda_{M}}{\lambda_{M}}$, $\frac{\delta \lambda_{B}}{\lambda_{B}}$ and $\frac{\delta G_{PMB}}{G_{PMB}}$ unambiguously. Even if we know the {\bf true values} of the $\lambda_P$, $\lambda_M$ and $\lambda_B$, setting the $\lambda_P$, $\lambda_M$ and $\lambda_B$ to be the true values results in unknown uncertainties, as the $\lambda_P$, $\lambda_M$ and $\lambda_B$ in the three-point QCD sum rules originate  from the QCD side, are not necessary to be the true values, and not to mention of the collective uncertainties $\delta\,\lambda_{P}\lambda_{M}\lambda_{B}G_{PMB}$.
On the other hand, we would overestimate the uncertainties, if we take the hadron masses  and pole residues as independent parameters, and endow they have their own uncertainties.
We have no choice but to take the approximation $\frac{\delta \lambda_{P}}{\lambda_{P}} =\frac{\delta \lambda_{M}}{\lambda_{M}}=\frac{\delta \lambda_{B}}{\lambda_{B}}=\frac{\delta G_{PMB}}{G_{PMB}}$ in Eq.\eqref{Uncertainty-4}, it is not a bad approximation. We fit the parameters $C_{P^{\prime}MB}$ to subtract contaminations and  produce flat Borel platforms,
the uncertainty $\delta m_c$ leads to a uncertainty $\pm 0.054\,\rm{GeV}$ for the integral intervals, such a large uncertainty have to be compensated by a huge change of the $C_{P^{\prime}MB}$, and is not under well control, we neglect the uncertainty $\delta m_c$.
 On the other hand, the uncertainties of the vacuum condensates cannot destroy the flatness, we do not need to add the terms $\delta C_{P^{\prime}MB}$, as such a lucky  happens  and we would like to set  $\delta C_{P^{\prime}MB}$ to be zero for simplicity.
In the two-point QCD sum rules, we often vary the  continuum threshold parameters $s_0$ to reproduce the central values of the experimental masses of the traditional mesons and baryons. The uncertainties $\delta s_0$ also lead to uncertainties $\delta m_{M}$ and $\delta m_{B}$,   we can  choose the experimental masses and  central values of the $s_0$ for simplicity and to avoid double-counting. When precise experimental data are accumulated, we can perform
a comprehensive and detailed error analysis, and obtain strict constraints on the input parameters, the present calculations serve as a preliminary estimation.

Then we obtain the uncertainties,
\begin{eqnarray}
\delta\,\lambda_{P}\lambda_{M}\lambda_{B}G_{PMB} &=&\lambda_{P}\lambda_{M}\lambda_{B}G_{PMB} \frac{4\delta G_{PMB}}{G_{PMB}}\, .
\end{eqnarray}

In calculations, we observe that it is better to choose  the free parameters  as
\begin{eqnarray}\label{C-fitted}
C_{P^{\prime}\eta_c\Lambda}&=&-1.78\times 10^{-5}{\rm{GeV}^9}-5.27\times 10^{-7}\,{\rm{GeV}^7}\,T^2\, ,\nonumber\\
C_{P^{\prime}\bar{D}_s\Lambda}&=&-5.51\times 10^{-6}{\rm{GeV}^9}-3.85\times 10^{-7}\,{\rm{GeV}^7}\,T^2\, ,\nonumber\\
C_{V}&=&2.52\times 10^{-5}{\rm{GeV}^9}+7.53\times 10^{-7}\,{\rm{GeV}^7}\,T^2\, ,\nonumber\\
C_{T}&=&7.00\times 10^{-7}{\rm{GeV}^{9}}+2.40\times 10^{-7}\,{\rm{GeV}^7}\,T^2\, ,
\end{eqnarray}
  to obtain flat platforms in the Borel windows   $T^2 =(8.9-9.9)\,\rm{GeV^2}$,
   $(4.6-5.6)\,\rm{GeV^2}$,   $(9.0-10.0)\,\rm{GeV^2}$ and
   $(9.0-10.0)\,\rm{GeV^2}$ for the hadronic coupling constants $G_{P\eta_c\Lambda}$, $G_{P\bar{D}_s\Lambda_c}$,  $G_V$ and $G_T$, respectively. We fit these values, see Eq.\eqref{C-fitted},  to obtain the same intervals  $T_{max}^2-T^2_{min}=1.0\,\rm{GeV}^2$ as in previous works \cite{Duality-WangZG-ZhangJX-EPJC-2018,Duality-WangZG-EPJC-2019,
Duality-Penta-mole-WangZG-CPC-2020,Duality-Penta-WangZG-CPC-2021,
Duality-WangZG-CPC-2022,Duality-WangZG-CPC-2022-2,Duality-WangZG-PRD-2024,
Duality-WangZG-AAPPS-2024,Duality-YangXS-CPC-2025,
Duality-WangZG-NPB-2024,Duality-YangXS-CPC-2026}, where the $T^2_{max}$ and $T^2_{min}$ are  the maximum and minimum values of the Borel windows, respectively. All in all, in the QCD sum rules in Eqs.\eqref{QCDSR-etacN}-\eqref{QCDSR-GV-GT}, we obtain the Borel parameters $T_{P\eta_c\Lambda}^2\approx T_{PJ/\psi\Lambda}^2=\mathcal{O}(m_{J/\psi}^2)$ and
$T^2_{P\bar{D}_s\Lambda_c}=\mathcal{O}(m_{\Lambda_c}^2)$, and the weight functions
$\exp\left(-\frac{m_{\eta_c}^2}{T_{P\eta_c\Lambda}^2}\right)\approx
\exp\left(-\frac{m_{J/\psi}^2}{T^2_{PJ/\psi\Lambda}}\right)\approx \exp\left(-\frac{m_{\Lambda_c}^2}{T^2_{P\bar{D}_s\Lambda_c}}\right)\approx e^{-1}$, the higher resonances and continuum states are depressed sufficiently indeed. The $C_{P^{\prime}MB}$ receive contributions from the higher resonances in the pentaquark channel (see Eqs.\eqref{define-C-etacLambda}-\eqref{define-C-JpsiLambda}), direct couplings  of the current $J_P(0)$ to the baryon-meson pairs (see Eq.\eqref{C-BM-pair}), and the 1P excitations in the $\Lambda$ and $\Lambda_c$ channels (see Eqs.\eqref{1P-etacLambda}-\eqref{1P-JpsiLambda}), in the  Borel windows, their total  contributions  account for  about $94\%$, $82\%$, $94\%$ and $74\%$  for the QCD sum rules in Eqs.\eqref{QCDSR-etacN}-\eqref{QCDSR-GV-GT} sequentially.

In calculations, we observe that it is the ultimate limit to obtain the Borel windows as large as $T_{max}^2-T^2_{min}=1.0\,\rm{GeV}^2$ via numerous trial and error. Before (after) reaching the ultimate limit, the predicted hadronic coupling constants $G$ decrease significantly  with larger (smaller) interval $T_{max}^2-T^2_{min}$. For example, in the case of the  $G_{P\eta_c\Lambda}$, if we accept a smaller  interval,  $T_{max}^2-T^2_{min}=0.8\,\rm{GeV}^2$, we could obtain the central value as large as $0.40$ (compared to the central value $0.30$ in Eq.\eqref{coupling-value}).  In fact, in most cases of calculating the hadronic coupling constants of the multiquark states, we could not obtain Borel windows as large as $1\,\rm{GeV}^2$, and we  gave up halfway.

\begin{figure}
\centering
\includegraphics[totalheight=6cm,width=7cm]{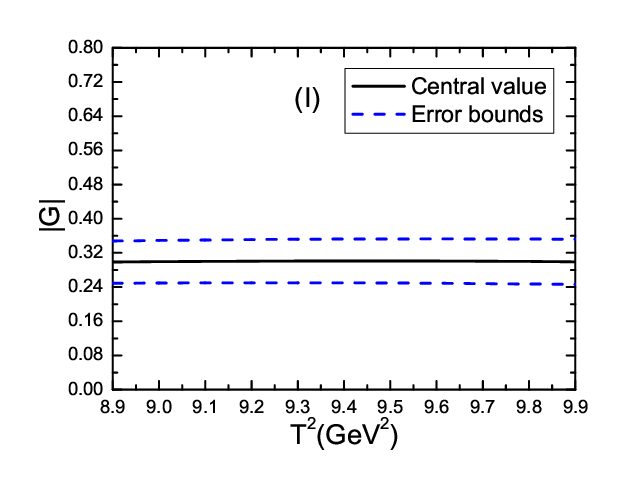}
\includegraphics[totalheight=6cm,width=7cm]{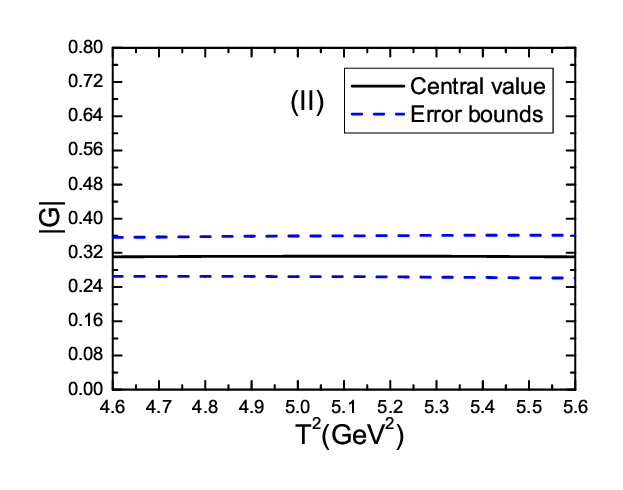}
\includegraphics[totalheight=6cm,width=7cm]{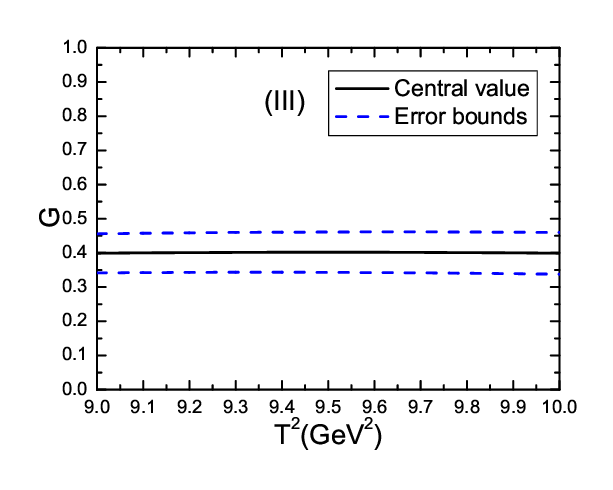}
\includegraphics[totalheight=6cm,width=7cm]{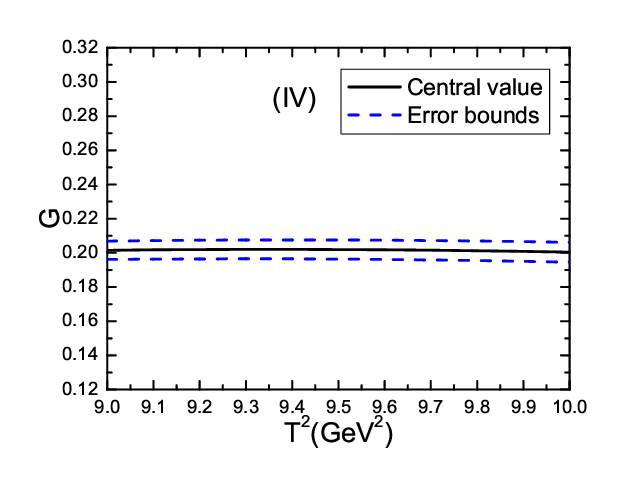}
  \caption{ The hadronic coupling constants   with variations of the  Borel parameters $T^2$, where the (I), (II), (III) and (IV) correspond to the $G_{P\eta_c\Lambda}$, $G_{P\bar{D}_s\Lambda_c}$,  $G_V$ and $G_T$, respectively. }\label{coupling-constants-fig}
\end{figure}

At last, we take  account of  the uncertainties  of the relevant    parameters (including the Borel parameters) on the QCD side,
and obtain the values of  the hadronic coupling constants $G_{P\eta_c\Lambda}$, $G_{P\bar{D}_s\Lambda_c}$,  $G_V$ and $G_T$, which are shown clearly in Fig.\ref{coupling-constants-fig},
\begin{eqnarray}\label{coupling-value}
|G_{P\eta_c\Lambda}|&=&0.30\pm0.05 \, , \nonumber\\
|G_{P\bar{D}_s\Lambda_c}|&=&0.31\pm0.05\, ,\nonumber\\
G_V&=&0.40\pm0.06 \, ,\nonumber\\
G_T&=&0.20\pm0.01 \, .
\end{eqnarray}
It is direct to  calculate the partial decay widths of the two-body strong decays,
\begin{eqnarray}
\Gamma\left(P_{cs}(4338)\to \eta_c \Lambda\right)
&=&\frac{p_f(m_P,m_{\eta_c},m_{\Lambda})}{8\pi m_P^2}\left[\left(m_P+m_{\Lambda}\right)^2-m^2_{\eta_c} \right] G_{P\eta_c\Lambda}^2\, , \nonumber\\
&=&27.6715 \, G_{P\eta_c\Lambda}^2 \,\,{\rm{MeV}}\,\, (28.2003  \, G_{P\eta_c\Lambda}^2 \,\,{\rm{MeV}})\, , \nonumber\\
&=&2.49\pm0.83\,{\rm{MeV}}\,\,(2.21\pm0.74\,{\rm{MeV}})\, , \nonumber\\
\Gamma\left(P_{cs}(4338)\to \bar{D}_s \Lambda_c\right)
&=&\frac{p_f(m_P,m_{D_s},m_{\Lambda_c})}{8\pi m_P^2}\left[\left(m_P+m_{\Lambda_c}\right)^2-m^2_{D_s} \right] G_{P\bar{D}_s\Lambda_c}^2\, , \nonumber\\
&=&33.7784 \, G_{P\bar{D}_s\Lambda_c}^2 \,{\rm{MeV}}\,\,(35.5673 \, G_{P\bar{D}_s\Lambda_c}^2 \,{\rm{MeV}})\, ,\nonumber\\
&=&3.25\pm1.05\,{\rm{MeV}}\,\,(2.74\pm 0.88\,{\rm{MeV}})\, , \nonumber\\
\Gamma\left(P_{cs}(4338)\to J/\psi \Lambda\right)&=&\frac{p_f(m_P,m_{J/\psi},m_{\Lambda})}{16\pi m_P^2}\left(C_{TT} \, G_T^2 + C_{TV}\,G_V G_T + C_{VV}\, G_V^2\right)\, ,\nonumber\\
&=&19.3429 \, G_T^2 - 67.1466\,G_V G_T + 58.3447\, G_V^2\,\,{\rm MeV}\, ,\nonumber\\
&&(20.0492 \, G_T^2 - 69.6383\,G_V G_T + 60.5548\, G_V^2\,\,{\rm MeV})\, ,\nonumber\\
&=&4.74\pm2.00\,\,{\rm{MeV}}\,\,(3.95\pm1.67\,\,{\rm{MeV}})\, ,
\end{eqnarray}
where
\begin{eqnarray}
C_{TT}&=&\frac{4(m^2_P -m^2_\Lambda)^2
+4m_{J/\psi}^2(m^2_\Lambda+m^2_P)-2m^4_{J/\psi}
-6m_{J/\psi}^2(m_P-m_{\Lambda})^2}{\left(m_P+m_{\Lambda}\right)^2}\, ,\nonumber\\
C_{TV}&=&\frac{12\left(m_{\Lambda}-m_P\right)
\left[(m_\Lambda+m_P)^2-m_{J/\psi}^2\right]
}{m_P+m_{\Lambda}}\, ,\nonumber\\
C_{VV}&=&\frac{2(m^2_P -m^2_\Lambda)^2-4m^4_{J/\psi}-4m_{J/\psi}^2(m^2_\Lambda+m^2_P)
+6m_{J/\psi}^2(m_P+m_{\Lambda})^2
}{m_{J/\psi}^2}\, ,
\end{eqnarray}
 $p_f(a,b,c)=\sqrt{\lambda(a^2,b^2,c^2)}/2a$, $\lambda(a,b,c)=a^2+b^2+c^2-2ab-2bc-2ca$.
In the brackets, we present the partial decay widths by choosing the experimental mass of the $P_{cs}(4338)$ and taking  account of the factors shown in Eqs.\eqref{Factor-Gamma}-\eqref{Factor-ms-decay}. Thereafter, we would like to present the decay width and ratios under such conditions in the brackets.

We calculate the uncertainty $\delta f$ using the following formula as usual,
\begin{eqnarray}
\delta f&=&\sqrt{\sum_{i} \left(\frac{\partial f(x)}{\partial x_i}\delta x_i\right)^2}\, ,
\end{eqnarray}
where the $f(x)$ denotes the total decay width, the $x_i$ denotes the partial decay widths.
If we saturate the total decay width of the $P_{cs}(4338)$ with the two-body strong decays  to the final states $\eta_c \Lambda$, $\bar{D}_s \Lambda_c$ and $J/\psi \Lambda$, we obtain the total width
$\Gamma(P_{cs}(4338))=10.48\pm 2.41\,\rm{MeV}$ ($8.90\pm 2.03\,\rm{MeV}$), which is consistent  with the experimental data  $\Gamma(P_{cs}(4338)) = 7.0\pm1.2\pm1.3 \mbox{ MeV}$ from the LHCb collaboration \cite{LHCb-Pcs4338-2023}. If we insist on that the uncertainties of the  hadronic coupling constants originate from the uncertainties of the  same input parameters, and add the uncertainties linearly, we obtain the uncertainty $\delta \Gamma(P_{cs}(4338))=\pm 3.88\,\rm{MeV}$ ($\pm 3.29\,\rm{MeV}$).
 It is reasonable to assign the
$P_{cs}(4338)$ as a $[us][dc]\bar{c}-[ds][uc]\bar{c}$ type pentaquark state with  the spin-parity  $J^P={\frac{1}{2}}^-$. The $Q$-value in the two-body strong decay $P_{cs}(4338)\to \bar{D}\Xi_c$ ($\bar{D}_s\Lambda_c$) is about $1\,\rm{MeV}$ ($83\,\rm{MeV}$), it is reasonable to neglect the channel $P_{cs}(4338)\to \bar{D}\Xi_c$.

In the scenario of molecular states, the total decay width of the $P_{cs}(4338)$  is saturated by the two channels $P_{cs}(4338)\to J/\psi\Lambda$ and $\eta_c \Lambda$. In Ref.\cite{Pcs4338-mole-decay-WangXW-PRD}, we
 obtained the
 total width $\Gamma(P_{cs}(4338))=6.16^{+8.01}_{-5.23}\,\rm{MeV}$ and the central ratio,
  \begin{eqnarray}
\frac{\Gamma\left(P_{cs}(4338)\to \eta_c\Lambda\right)}{\Gamma\left(P_{cs}(4338)\to J/\psi\Lambda\right)}&=&0.18\, .
\end{eqnarray}
In Ref.\cite{Pcs4338-mole-decay-Azizi-PRD-2023},
 Azizi, Sarac and  Sundu only took account of the ground state contributions  in all channels ($P_{cs}(4338)$, $\Lambda$, $J/\psi$, $\eta_c$) on the hadron side, then performed double Borel transformation in regard  to the $P_{cs}(4338)$ and $\Lambda$ channels  respectively, and parameterized the hadronic coupling constants with the $\eta_c(J/\psi)$ channel off-shell with some model functions. At last,  they extracted the $Q^2 $ to the physical regions $-m_{\eta_c}^2$ and $-m^2_{J/\psi}$.
 They obtained the total width $\Gamma(P_{cs}(4338))=10.40\pm 1.93\,\rm{MeV}$ and the central ratio,
  \begin{eqnarray}
\frac{\Gamma\left(P_{cs}(4338)\to \eta_c\Lambda\right)}{\Gamma\left(P_{cs}(4338)\to J/\psi\Lambda\right)}&=&0.44\, .
\end{eqnarray}
In the present work, if we also saturate the total decay  width of the $P_{cs}(4338)$  with the two channels $P_{cs}(4338)\to J/\psi\Lambda$ and $\eta_c \Lambda$, we obtain the  total width $\Gamma(P_{cs}(4338))=7.23\pm 2.17\,\rm{MeV}$ ($6.16\pm 1.83\,\rm{MeV}$) and the  ratio,
  \begin{eqnarray}
\frac{\Gamma\left(P_{cs}(4338)\to \eta_c\Lambda\right)}{\Gamma\left(P_{cs}(4338)\to J/\psi\Lambda\right)}&=&0.53 \pm 0.28 \, \, (0.56\pm0.30)\, .
\end{eqnarray}
In Ref.\cite{Pcs4338-Braaten-BO},  Alasiri, Braaten and Bruschini
assigned the  $P_{cs}(4338)$  as a color octet-octet type  $c\bar{c}$-$uds$
state in the Born-Oppenheimer  approximation, and obtained  the ratio,
\begin{eqnarray}
\frac{\Gamma\left(P_{cs}(4338)\to \eta_c\Lambda\right)}{\Gamma\left(P_{cs}(4338)\to J/\psi\Lambda\right)}&=&0.43\, .
\end{eqnarray}
No definitive  conclusion could be obtained at the present time, as neglecting  the decay channel $P_{cs}(4338)\to \bar{D}_s\Lambda_c$ leads to unknown uncertainty  in the molecule scenario \cite{Pcs4338-mole-decay-WangXW-PRD,Pcs4338-mole-decay-Azizi-PRD-2023}, this partial decay width is not necessary to be neglectful.  Furthermore, we obtain the ratio,
\begin{eqnarray}
\frac{\Gamma\left(P_{cs}(4338)\to \bar{D}_s\Lambda_c\right)}{\Gamma\left(P_{cs}(4338)\to J/\psi\Lambda\right)}&=&0.69\pm0.36\,\,(0.69\pm 0.37)\, ,
\end{eqnarray}
 the partial decay width $\Gamma(\bar{D}_s\Lambda_c)$ should be taken into account.
Precisely  measuring the total width and branching fractions  is still needed to distinguish the pentaquark and molecular scenarios.

\section{Conclusion}
In this work, we identify the $P_{cs}(4338)$ as the $[us][dc]\bar{c}-[ds][uc]\bar{c}$ type  pentaquark state with  the spin-parity $J^P={\frac{1}{2}}^-$, and study its  hadronic coupling constants via the three-point QCD sum rules by adopting  rigorous  quark-hadron duality introduced in our previous work for the first time. We select  the optimal parametrization to subtract  the contaminations from the higher resonances and continuum states involving  the pentaquark channel (also the baryon channels with negative parity, and direct couplings to the baryon-meson pairs), then fit the optimal  values to obtain flat platforms in the Borel windows, therefore extract  reliable    hadronic coupling constants. At last,  we obtain the total width $\Gamma(P_{cs}(4338))=10.48\pm 2.41\,\rm{MeV}$ ($8.90\pm 2.03\,\rm{MeV}$), which is consistent  with the experimental data  $\Gamma(P_{cs}(4338)) = 7.0\pm1.2\pm1.3 \mbox{ MeV}$ from the LHCb collaboration within uncertainties, and supports identifying  the $P_{cs}(4338)$ as the  $[us][dc]\bar{c}-[ds][uc]\bar{c}$  type compact pentaquark state with the spin-parity  $J^P={\frac{1}{2}}^-$. In the molecule scenario, the decay channel $P_{cs}(4338)\to \bar{D}_s\Lambda_c$ is neglected, which leads to unknown uncertainty, as this partial decay width is not necessary to be neglectful.
The predicted  partial decay widths  can be compared to the experimental data   in the future.

\section*{Appendix}
The explicit expressions of the  spectral densities $\rho^{\eta_c\Lambda}_{QCD}(s,u)$, $\rho^{\bar{D}_s\Lambda_c}_{QCD}(s,u)$
and $\rho^{J/\psi \Lambda,1/2}_{QCD}(s,u)$,
\begin{eqnarray}
\rho^{\eta_c\Lambda}_{QCD}(s,u)&=&\sum_{i}\rho^{\eta_c\Lambda}_{i}(s,u)\, ,\nonumber\\
\rho^{\bar{D}_s\Lambda_c}_{QCD}(s,u)&=&\sum_{i}\rho^{\bar{D}_s\Lambda_c}_{i}(s,u)\, ,\nonumber\\
\rho^{J/\psi\Lambda,1/2}_{QCD}(s,u)&=&\sum_{i}\rho^{J/\psi\Lambda}_{1/2,i}(s,u)\, ,
\end{eqnarray}
where $i=0$, $3$, $\cdots$, $10$,

\begin{eqnarray}\label{etacLambda-0}
\rho^{\eta_c\Lambda}_{0}(s,u)
&=& -\frac{3m_c}{2048 \pi^6} \int_{x_i}^{x_f} dx\, u^2\, ,
\end{eqnarray}

\begin{eqnarray}
\rho^{\eta_c\Lambda}_{3}(s,u)
&=& \frac{m_sm_c\left[4\langle\bar{q}q\rangle-3 \langle\bar{s}s\rangle\right]}{128\pi^4} \int_{x_i}^{x_f} dx\, ,
\end{eqnarray}

\begin{eqnarray}
\rho^{\eta_c\Lambda}_{4}(s,u)
&=& \frac{m_c^3\langle \tilde{\alpha}_s GG \rangle}{6144\pi^4 T^4} \int_{x_i}^{x_f} dx\, \frac{1}{x^3}\, u^2\, \delta(s-\tilde{m}_c^2) \nonumber\\
&&-\frac{m_c\langle \tilde{\alpha}_s GG \rangle}{2048\pi^4 T^2} \int_{x_i}^{x_f} dx\, \frac{1-x}{x^2}\, u^2\, \delta(s-\tilde{m}_c^2)-\frac{3m_c\langle \tilde{\alpha}_s GG \rangle}{1024\pi^4} \int_{x_i}^{x_f} dx \nonumber\\
&&  +\frac{m_c\langle \tilde{\alpha}_s GG \rangle}{6144\pi^4} \int_{x_i}^{x_f} dx\, \frac{1}{x(1-x)}\, u \left(1-\frac{3u}{2T^2} \right)\,\delta(s-\tilde{m}_c^2) \nonumber\\
&& -\frac{m_s\langle \tilde{\alpha}_s GG \rangle}{36864\pi^4} \int_{x_i}^{x_f}dx\, u \,\delta(s-\tilde{m}_c^2)\, ,
\end{eqnarray}

\begin{eqnarray}
\rho^{\eta_c\Lambda}_{5}(s,u)
&=& -\frac{m_sm_c\left[\langle \bar{q}G q \rangle - \langle \bar{s}G s \rangle\right]}{128\pi^4} \int_{x_i}^{x_f} dx\, \delta(u)  -\frac{7m_s m_c\langle \bar{q}G q \rangle}{3072\pi^4} \int_{x_i}^{x_f} dx\, \frac{1}{x} \,\delta(s-\tilde{m}_c^2) \nonumber\\
&& +\frac{5\langle \bar{q}G q \rangle + \langle \bar{s}G s \rangle}{9216\pi^4} \int_{x_i}^{x_f} dx\,  u \,\delta(s-\tilde{m}_c^2)\, ,
\end{eqnarray}

\begin{eqnarray}
\rho^{\eta_c\Lambda}_{6}(s,u)
&=& -\frac{m_c\left[4\langle\bar{q}q\rangle\langle\bar{s}s\rangle-\langle\bar{q}q\rangle^2\right]}{48\pi^2} \int_{x_i}^{x_f} dx\, \delta(u)\, ,
\end{eqnarray}

\begin{eqnarray}
\rho^{\eta_c\Lambda}_{7}(s,u)
&=& -\frac{m_sm_c^3\left[4\langle\bar{q}q\rangle-3\langle\bar{s}s\rangle\right]
\langle \tilde{\alpha}_s GG \rangle}{1152\pi^2 T^4} \int_{x_i}^{x_f} dx\, \frac{1}{x^3}\, \delta(s-\tilde{m}_c^2) \nonumber\\
&& +\frac{m_sm_c\left[4\langle\bar{q}q\rangle-3\langle\bar{s}s\rangle\right]\langle \tilde{\alpha}_s GG \rangle}{384\pi^2 T^2} \int_{x_i}^{x_f} dx\, \frac{1-x}{x^2} \,\delta(s-\tilde{m}_c^2) \nonumber\\
&& +\frac{m_sm_c\left[\langle\bar{s}s\rangle-5\langle\bar{q}q\rangle\right]\langle \tilde{\alpha}_s GG \rangle}{2304\pi^2} \int_{x_i}^{x_f} dx\, \frac{1}{x}\, \delta(s-\tilde{m}_c^2)\,\delta(u) \nonumber\\
&& +\frac{m_sm_c\left[4\langle\bar{q}q\rangle-3\langle\bar{s}s\rangle\right]\langle \tilde{\alpha}_s GG \rangle}{768\pi^2 T^2} \int_{x_i}^{x_f} dx\, \frac{1}{x(1-x)}\, \delta(s-\tilde{m}_c^2) \nonumber\\
&& -\frac{m_s m_c \left[3\langle\bar{s}s\rangle-2\langle\bar{q}q\rangle\right]\langle \tilde{\alpha}_s GG \rangle}{1152\pi^2 T^2} \int_{x_i}^{x_f} dx\, \delta(u) \, ,
\end{eqnarray}

\begin{eqnarray}
\rho^{\eta_c\Lambda}_{8}(s,u)
&=& \frac{m_c\left[2\langle\bar{q}q\rangle\langle \bar{s} G s \rangle + 2\langle\bar{s}s\rangle\langle \bar{q}G q \rangle - \langle\bar{q}q\rangle\langle \bar{q} G q \rangle\right]}{192\pi^2 T^2} \int_{x_i}^{x_f} dx\, \delta(u) \nonumber\\
&& +\frac{m_c\left[4\langle\bar{q}q\rangle\langle \bar{s} G s \rangle + 7\langle\bar{s}s\rangle\langle \bar{q}G q \rangle - 5\langle\bar{q}q\rangle\langle \bar{q} G q \rangle\right]}{1152\pi^2} \int_{x_i}^{x_f} dx\, \frac{1}{x}\, \delta(s-\tilde{m}_c^2)\,\delta(u) \nonumber\\
&& -\frac{5m_s\langle\bar{s}s\rangle\langle \bar{q}G q \rangle}{2304\pi^2} \int_{x_i}^{x_f} dx\,   \delta(s-\tilde{m}_c^2)\,\delta(u) \, ,
\end{eqnarray}

\begin{eqnarray}
\rho^{\eta_c\Lambda}_{10}(s,u)
&=& \frac{m_c^3\left[4\langle\bar{q}q\rangle\langle\bar{s}s\rangle-\langle\bar{q}q\rangle^2\right]\langle \tilde{\alpha}_s GG \rangle}{432 T^4} \int_{x_i}^{x_f} dx\, \frac{1}{x^3} \, \delta(s-\tilde{m}_c^2)\,\delta(u) \nonumber\\
&& -\frac{m_c\left[4\langle\bar{q}q\rangle\langle\bar{s}s\rangle-\langle\bar{q}q\rangle^2\right]\langle \tilde{\alpha}_s GG \rangle}{144 T^2} \int_{x_i}^{x_f} dx\, \frac{1-x}{x^2}\,  \delta(s-\tilde{m}_c^2)\,\delta(u) \nonumber\\
&& +\frac{m_c\left[13\langle\bar{q}q\rangle^2-43\langle\bar{q}q\rangle\langle\bar{s}s\rangle\right]\langle \tilde{\alpha}_s GG \rangle}{3456 T^2} \int_{x_i}^{x_f} dx\, \frac{1}{x(1-x)}\, \delta(s-\tilde{m}_c^2)\,\delta(u) \nonumber\\
&& -\frac{m_s\langle\bar{q}q\rangle\langle\bar{s}s\rangle\langle \tilde{\alpha}_s GG \rangle}{3456 T^2} \int_{x_i}^{x_f} dx\, \delta(s-\tilde{m}_c^2)\,\delta(u) \nonumber\\
&& +\frac{5m_s\langle \bar{q}Gq \rangle\langle \bar{s} G s \rangle}{6912\pi^2 T^2} \int_{x_i}^{x_f} dx\,  \delta(s-\tilde{m}_c^2)\,\delta(u) \nonumber\\
&& -\frac{m_c\left[20\langle \bar{q}Gq \rangle^2+\langle \bar{q}Gq \rangle\langle \bar{s}Gs \rangle\right]}{18432\pi^2 T^2} \int_{x_i}^{x_f}  dx\, \frac{1}{x(1-x)} \,\delta(s-\tilde{m}_c^2)\,\delta(u) \, ,
\end{eqnarray}

\begin{eqnarray}
\rho^{J/\psi\Lambda}_{1,0}(s,u)
&=& -\frac{3m_c}{2048 \pi^6} \int_{x_i}^{x_f} dx\, u^2\, ,
\end{eqnarray}

\begin{eqnarray}
\rho^{J/\psi\Lambda}_{1,3}(s,u)
&=& \frac{m_sm_c\left[4\langle\bar{q}q\rangle-3 \langle\bar{s}s\rangle\right]}{128\pi^4} \int_{x_i}^{x_f} dx \, ,
\end{eqnarray}

\begin{eqnarray}
\rho^{J/\psi\Lambda}_{1,4}(s,u)
&=& \frac{m_c^3\langle \tilde{\alpha}_s GG \rangle}{6144\pi^4 T^4} \int_{x_i}^{x_f} dx\, \frac{1}{x^3} \, u^2\, \delta(s-\tilde{m}_c^2)  -\frac{m_c\langle \tilde{\alpha}_s GG \rangle}{2048\pi^4 T^2} \int_{x_i}^{x_f} dx\, \frac{1-x}{x^2} \, u^2 \, \delta(s-\tilde{m}_c^2) \nonumber\\
&& -\frac{3m_c\langle \tilde{\alpha}_s GG \rangle}{1024\pi^4} \int_{x_i}^{x_f} dx +\frac{m_c\langle \tilde{\alpha}_s GG \rangle}{6144\pi^4} \int_{x_i}^{x_f} dx\, \frac{1}{x(1-x)} \, u\left(1+\frac{u}{2T^2}\right) \delta(s-\tilde{m}_c^2) \nonumber\\
&& -\frac{m_s\langle \tilde{\alpha}_s GG \rangle}{73728\pi^4} \int_{x_i}^{x_f} dx\,  u\, \delta(s-\tilde{m}_c^2)\, ,
\end{eqnarray}

\begin{eqnarray}
\rho^{J/\psi\Lambda}_{1,5}(s,u)
&=& -\frac{m_sm_c\left[\langle \bar{q}G q \rangle - \langle \bar{s}G s \rangle\right]}{128\pi^4} \int_{x_i}^{x_f} dx\, \delta(u)  -\frac{m_sm_c \langle \bar{q}G q \rangle}{768\pi^4} \int_{x_i}^{x_f} dx\, \frac{1}{x}\, \delta(s-\tilde{m}_c^2) \nonumber\\
&& +\frac{m_s m_c \langle \bar{q}G q \rangle}{1024\pi^4} \int_{x_i}^{x_f} dx\, \delta(s-\tilde{m}_c^2)  +\frac{5\langle \bar{q}G q \rangle + \langle \bar{s}G s \rangle}{18432\pi^4} \int_{x_i}^{x_f} dx\,  u\, \delta(s-\tilde{m}_c^2)\, ,
\end{eqnarray}

\begin{eqnarray}
\rho^{J/\psi\Lambda}_{1,6}(s,u)
&=& -\frac{m_c\left[4\langle\bar{q}q\rangle\langle\bar{s}s\rangle-\langle\bar{q}q\rangle^2\right]}{48\pi^2} \int_{x_i}^{x_f} dx\, \delta(u) \, ,
\end{eqnarray}

\begin{eqnarray}
\rho^{J/\psi\Lambda}_{1,7}(s,u)
&=& -\frac{m_sm_c^3\left[4\langle\bar{q}q\rangle-3\langle\bar{s}s\rangle\right]\langle \tilde{\alpha}_s GG \rangle}{1152\pi^2 T^4} \int_{x_i}^{x_f} dx\, \frac{1}{x^3} \,\delta(s-\tilde{m}_c^2) \nonumber\\
&& +\frac{m_sm_c\left[4\langle\bar{q}q\rangle-3\langle\bar{s}s\rangle\right]\langle \tilde{\alpha}_s GG \rangle}{384\pi^2 T^2} \int_{x_i}^{x_f} dx\, \frac{1-x}{x^2} \,\delta(s-\tilde{m}_c^2) \nonumber\\
&& -\frac{m_sm_c \left[3\langle\bar{s}s\rangle-2\langle\bar{q}q\rangle\right]\langle \tilde{\alpha}_s GG \rangle}{1152\pi^2 T^2} \int_{x_i}^{x_f} dx\, \delta(u) \nonumber\\
&& +\frac{m_sm_c\left[\langle\bar{s}s\rangle-5\langle\bar{q}q\rangle\right]\langle \tilde{\alpha}_s GG \rangle}{4608\pi^2} \int_{x_i}^{x_f} dx\, \frac{1}{x(1-x)} \, \delta(s-\tilde{m}_c^2)\,\delta(u) \nonumber\\
&& -\frac{m_sm_c\left[4\langle\bar{q}q\rangle-3\langle\bar{s}s\rangle\right]\langle \tilde{\alpha}_s GG \rangle}{2304\pi^2 T^2} \int_{x_i}^{x_f} dx\, \frac{1}{x(1-x)}\, \delta(s-\tilde{m}_c^2) \, ,
\end{eqnarray}

\begin{eqnarray}
\rho^{J/\psi\Lambda}_{1,8}(s,u)
&=& \frac{m_c\left[2\langle\bar{q}q\rangle\langle \bar{s}Gs \rangle+2\langle\bar{s}s\rangle\langle \bar{q}Gq \rangle-\langle\bar{q}q\rangle\langle \bar{q}Gq \rangle\right]}{192\pi^2 T^2} \int_{x_i}^{x_f} dx\, \delta(u) \nonumber\\
&& +\frac{m_c\left[2\langle\bar{q}q\rangle\langle \bar{s}Gs \rangle+2\langle\bar{s}s\rangle\langle \bar{q}Gq \rangle-\langle\bar{q}q\rangle\langle \bar{q}Gq \rangle\right]}{576\pi^2} \int_{x_i}^{x_f} dx\, \frac{1}{x}\, \delta(s-\tilde{m}_c^2)\,\delta(u) \nonumber\\
&& -\frac{m_c\left[\langle\bar{s}s\rangle\langle \bar{q}Gq \rangle-\langle\bar{q}q\rangle\langle \bar{q}Gq \rangle\right]}{384\pi^2} \int_{x_i}^{x_f} dx\, \delta(s-\tilde{m}_c^2)\, \delta(u) \nonumber\\
&& -\frac{5m_s\langle\bar{s}s\rangle\langle \bar{q}G q \rangle}{4608\pi^2} \int_{x_i}^{x_f} dx\,   \delta(s-\tilde{m}_c^2)\, \delta(u)\, ,
\end{eqnarray}

\begin{eqnarray}
\rho^{J/\psi\Lambda}_{1,10}(s,u)
&= & \frac{m_c^3\left[4\langle\bar{q}q\rangle\langle\bar{s}s\rangle-\langle\bar{q}q\rangle^2\right]\langle \tilde{\alpha}_s GG \rangle}{432 T^4} \int_{x_i}^{x_f} dx\, \frac{1}{x^3} \,\delta(s-\tilde{m}_c^2)\,\delta(u) \nonumber\\
&& -\frac{m_c\left[4\langle\bar{q}q\rangle\langle\bar{s}s\rangle-\langle\bar{q}q\rangle^2\right]\langle \tilde{\alpha}_s GG \rangle}{144 T^2} \int_{x_i}^{x_f} dx\, \frac{1-x}{x^2}\, \delta(s-\tilde{m}_c^2)\,\delta(u) \nonumber\\
&& -\frac{m_c\left[\langle\bar{q}q\rangle^2-7\langle\bar{q}q\rangle\langle\bar{s}s\rangle\right]\langle \tilde{\alpha}_s GG \rangle}{1152 T^2} \int_{x_i}^{x_f} dx\, \frac{1}{x(1-x)}\, \delta(s-\tilde{m}_c^2)\,\delta(u) \nonumber\\
&& -\frac{m_s\langle\bar{q}q\rangle\langle\bar{s}s\rangle\langle \tilde{\alpha}_s GG \rangle}{6912 T^2} \int_{x_i}^{x_f} dx\, \delta(s-\tilde{m}_c^2)\,\delta(u) +\frac{5m_s\langle\bar{q}Gq\rangle\langle \bar{s}Gs \rangle}{13824\pi^2 T^2} \int_{x_i}^{x_f} dx\,   \delta(s-\tilde{m}_c^2)\,\delta(u) \nonumber\\
&& -\frac{m_c\left[\langle\bar{q}Gq\rangle^2-7\langle\bar{q}Gq\rangle
\langle\bar{s}Gs\rangle\right]}{4608\pi^2 T^2} \int_{x_i}^{x_f} dx\, \delta(s-\tilde{m}_c^2)\, \delta(u) \nonumber\\
&& -\frac{m_c\left[2\langle\bar{q}Gq\rangle^2-5\langle\bar{q}Gq\rangle\langle\bar{s}Gs\rangle\right]}{2304\pi^2 T^2} \int_{x_i}^{x_f} dx\, \frac{1-x}{x} \, \delta(s-\tilde{m}_c^2)\, \delta(u) \nonumber\\
&& +\frac{37m_c\left[\langle\bar{q}Gq\rangle^2+2\langle
\bar{q}Gq\rangle\langle\bar{s}Gs\rangle\right]}{55296\pi^2 T^2} \int_{x_i}^{x_f} dx\, \frac{1}{x(1-x)}\, \delta(s-\tilde{m}_c^2)\,\delta(u)\, ,
\end{eqnarray}

\begin{eqnarray}
\rho^{J/\psi\Lambda}_{2,0}(s,u)
&=& \frac{3}{1024 \pi^6} \int_{x_i}^{x_f} dx\, x(1-x)\, u^2\,s
\, ,
\end{eqnarray}

\begin{eqnarray}
\rho^{J/\psi\Lambda}_{2,3}(s,u)
&=& -\frac{m_s\left[4\langle\bar{q}q\rangle-3 \langle\bar{s}s\rangle\right]}{64\pi^4} \int_{x_i}^{x_f} dx\, x(1-x)\,s \, ,
\end{eqnarray}

\begin{eqnarray}
\rho^{J/\psi\Lambda}_{2,4}(s,u)
&=& \frac{m_c^2\langle \tilde{\alpha}_s GG \rangle }{6144\pi^4 T^2} \int_{x_i}^{x_f} dx\, \frac{1-x}{x^2}\, \left(1-\frac{s}{T^2}\right)\,u^2\,\delta(s-\tilde{m}_c^2) \nonumber\\
&& +\frac{m_c^2\langle \tilde{\alpha}_s GG \rangle}{12288\pi^4 T^2} \int_{x_i}^{x_f} dx\, \frac{1}{x^2}\, u^2\left(6-\frac{s}{T^2}\right)\,\delta(s-\tilde{m}_c^2) \nonumber\\
&& +\frac{3\langle \tilde{\alpha}_s GG \rangle}{512\pi^4} \int_{x_i}^{x_f} dx\, x(1-x)\, s  -\frac{\langle \tilde{\alpha}_s GG \rangle}{1536\pi^4} \int_{x_i}^{x_f} dx\, \, su\,\delta(s-\tilde{m}_c^2) \nonumber\\
&& -\frac{\langle \tilde{\alpha}_s GG \rangle}{12288\pi^4 } \int_{x_i}^{x_f} dx\, \left(2+\frac{s}{T^2} \right)\,u^2\,\delta(s-\tilde{m}_c^2)\, ,
\end{eqnarray}

\begin{eqnarray}
\rho^{J/\psi\Lambda}_{2,5}(s,u)
&=& \frac{m_s\left[\langle \bar{q}G q \rangle - \langle \bar{s}G s \rangle\right]}{64\pi^4} \int_{x_i}^{x_f} dx\, x(1-x)\,s\,\delta(u) \nonumber\\
&& +\frac{m_s \langle \bar{q}G q \rangle}{1536\pi^4} \int_{x_i}^{x_f} dx\, s\,\delta(s-\tilde{m}_c^2)\, ,
\end{eqnarray}

\begin{eqnarray}
\rho^{J/\psi\Lambda}_{2,6}(s,u)
&= & \frac{4\langle\bar{q}q\rangle\langle\bar{s}s\rangle-
\langle\bar{q}q\rangle^2}{24\pi^2} \int_{x_i}^{x_f} dx\, x(1-x)\,s\,\delta(u)\, ,
\end{eqnarray}

\begin{eqnarray}
\rho^{J/\psi\Lambda}_{2,7}(s,u)
&= & -\frac{m_sm_c^2\left[4\langle\bar{q}q\rangle-3\langle\bar{s}s\rangle\right]\langle \tilde{\alpha}_s GG \rangle }{1152\pi^2 T^2} \int_{x_i}^{x_f} dx\, \frac{1-x}{x^2}\,\left(1 - \frac{s}{T^2}\right)\,\delta(s-\tilde{m}_c^2) \nonumber\\
&& -\frac{m_sm_c^2\left[4\langle\bar{q}q\rangle-3\langle\bar{s}s\rangle\right]\langle \tilde{\alpha}_s GG \rangle}{384\pi^2 T^2} \int_{x_i}^{x_f} dx\, \frac{1}{x^2}\,\left(6-\frac{s}{T^2}\right)\,\delta(s-\tilde{m}_c^2) \nonumber\\
&& +\frac{m_s\langle\bar{s}s\rangle\langle \tilde{\alpha}_s GG \rangle}{192\pi^2 T^2} \int_{x_i}^{x_f} dx\, x(1-x)\,s\,\delta(u) \nonumber\\
&& -\frac{m_s\left[ \langle\bar{s}s\rangle-5 \langle\bar{q}q\rangle\right]\langle \tilde{\alpha}_s GG \rangle}{1152\pi^2} \int_{x_i}^{x_f} dx\, s\,\delta(s-\tilde{m}_c^2)\,\delta(u) \nonumber\\
&& +\frac{m_s\left[4\langle\bar{q}q\rangle-3\langle\bar{s}s\rangle\right]\langle \tilde{\alpha}_s GG \rangle}{2304\pi^2 } \int_{x_i}^{x_f} dx\,\left(2 + \frac{s}{T^2}\right)\,\delta(s-\tilde{m}_c^2) \nonumber\\
&& -\frac{m_s\langle\bar{q}q\rangle\langle \tilde{\alpha}_s GG \rangle}{576\pi^2 T^2} \int_{x_i}^{x_f} dx\, (2s-\tilde{m}_c^2)\,\delta(u) \nonumber\\
&& -\frac{m_s m_c^2\langle\bar{q}q\rangle\langle \tilde{\alpha}_s GG \rangle}{576\pi^2 T^2} \int_{x_i}^{x_f} dx\, \delta(u)\, ,
\end{eqnarray}

\begin{eqnarray}
\rho^{J/\psi\Lambda}_{2,8}(s,u)
&=& -\frac{2\langle\bar{q}q\rangle\langle \bar{s}Gs \rangle+2\langle\bar{s}s\rangle\langle \bar{q}Gq \rangle-\langle\bar{q}q\rangle\langle \bar{q}Gq \rangle}{96\pi^2 T^2} \int_{x_i}^{x_f} dx\, x(1-x)\,s\,\delta(u) \nonumber\\
&& -\frac{ 4\langle\bar{q}q\rangle\langle \bar{s}Gs \rangle+\langle\bar{s}s\rangle\langle \bar{q}Gq \rangle+\langle\bar{q}q\rangle\langle \bar{q}Gq \rangle}{576\pi^2} \int_{x_i}^{x_f} dx\, s\,\delta(s-\tilde{m}_c^2)\,\delta(u)\, ,
\end{eqnarray}

\begin{eqnarray}\label{JpsiLambda-10}
\rho^{J/\psi\Lambda}_{2,10}(s,u)
&=& \frac{m_c^2\left[4\langle\bar{q}q\rangle\langle\bar{s}s\rangle-\langle\bar{q}q\rangle^2\right]\langle \tilde{\alpha}_s GG \rangle }{432 T^2} \int_{x_i}^{x_f} dx\, \frac{1-x}{x^2}\,\left(1 -\frac{s}{T^2} \right)\,\delta(s-\tilde{m}_c^2)\,\delta(u) \nonumber\\
&& +\frac{m_c^2\left[4\langle\bar{q}q\rangle\langle\bar{s}s\rangle
-\langle\bar{q}q\rangle^2\right]\langle \tilde{\alpha}_s GG \rangle}{864 T^2} \int_{x_i}^{x_f} dx\, \frac{1}{x^2}\,\left(6-\frac{s}{T^2}\right)\,\delta(s-\tilde{m}_c^2)\,\delta(u) \nonumber\\
&& -\frac{\left[\langle\bar{q}q\rangle^2+5\langle\bar{q}q\rangle\langle\bar{s}s\rangle\right]\langle \tilde{\alpha}_s GG \rangle }{864 T^2} \int_{x_i}^{x_f} dx\,s\, \delta(s-\tilde{m}_c^2)\,\delta(u) \nonumber\\
&& +\frac{\left[\langle\bar{q}q\rangle^2-4\langle\bar{q}q\rangle\langle\bar{s}s\rangle\right]\langle \tilde{\alpha}_s GG \rangle}{864 } \int_{x_i}^{x_f} dx\,
\left(2 +\frac{s}{T^2} \right)\,\delta(s-\tilde{m}_c^2)\,\delta(u) \nonumber\\
&& +\frac{8\langle\bar{q}Gq\rangle^2-29\langle\bar{q}Gq\rangle
\langle\bar{s}Gs\rangle}{27648\pi^2} \int_{x_i}^{x_f} dx\,\delta(s-\tilde{m}_c^2)\,\delta(u) \nonumber\\
&& -\frac{229\langle\bar{q}Gq\rangle^2+296\langle\bar{q}Gq\rangle
\langle\bar{s}Gs\rangle}{55296\pi^2 T^2} \int_{x_i}^{x_f} dx\, s\,\delta(s-\tilde{m}_c^2)\,\delta(u) \, ,
\end{eqnarray}

\begin{eqnarray}\label{DsLambdac-0}
\rho^{\bar{D}_s\Lambda_c}_{0}(s,u)
&=& -\frac{3m_c}{1024 \pi^6} \int_{x_i}^{1} dx\int_{y_i}^{1} dy\, (1-x)\,y(1-y)^2\,(u-\tilde{m}_y^2)^2 \nonumber\\
&& -\frac{3m_s}{1024 \pi^6} \int_{x_i}^{1} dx\int_{y_i}^{1} dy\, x\,y(1-y)^2\,(u-\tilde{m}_y^2)^2 \, ,
\end{eqnarray}

\begin{eqnarray}
\rho^{\bar{D}_s\Lambda_c}_{3}(s,u)
&=& \frac{\langle\bar{s}s\rangle}{256\pi^4}\left(1-\frac{m_s m_c}{2 T^2}\right) \int_{y_i}^{1} dy\, y(1-y)^2\,(u-\tilde{m}_y^2)^2\,\delta(s-m_c^2) \nonumber\\
&& +\frac{m_c^2\langle\bar{q}q\rangle}{64\pi^4} \int_{x_i}^{1} dx\int_{y_i}^{1} dy\, (1-x)\,(1-y) \nonumber\\
&& +\frac{m_s m_c\langle\bar{q}q\rangle}{64\pi^4} \int_{x_i}^{1} dx\int_{y_i}^{1} dy\, x\,(1-y)\, ,
\end{eqnarray}

\begin{eqnarray}
\rho^{\bar{D}_s\Lambda_c}_{4}(s,u)
&=& -\frac{m_c\langle \tilde{\alpha}_s GG \rangle}{6144\pi^4 T^2} \int_{x_i}^{1} dx\int_{y_i}^{1} dy\, \frac{y(1-x)(1-y)^2}{x^2}\left(3-\frac{s}{T^2}\right)(u-\tilde{m}_y^2)^2\,\delta(s-\tilde{m}_x^2) \nonumber\\
&& -\frac{\langle \tilde{\alpha}_s GG \rangle}{6144\pi^4 T^2} \int_{x_i}^{1} dx\int_{y_i}^{1} dy\, \frac{y(1-y)^2}{x}\left(3m_c-\frac{sm_s}{T^2}\right)\,(u-\tilde{m}_y^2)^2\,\delta(s-\tilde{m}_x^2) \nonumber\\
&& +\frac{ m_c^2\langle \tilde{\alpha}_s GG \rangle}{3072\pi^4} \int_{x_i}^{1} dx\int_{y_i}^{1} dy\, \frac{(1-y)^2}{y^2}\,\left[(1-x)m_c +x m_s\right]\,\delta(u-\tilde{m}_y^2) \nonumber\\
&& -\frac{\langle \tilde{\alpha}_s GG \rangle}{1024\pi^4} \int_{x_i}^{1} dx\int_{y_i}^{1} dy\, y\,\left[(1-x)m_c +x m_s\right]\nonumber\\
&& -\frac{m_c\langle \tilde{\alpha}_s GG \rangle}{12288\pi^4} \int_{x_i}^{1} dx\int_{y_i}^{1} dy\, \frac{(1-x)(1-y)^2}{x}\,(u-\tilde{m}_y^2)\,\delta(s-\tilde{m}_x^2) \nonumber\\
&& +\frac{\left(m_c+m_s\right)\langle \tilde{\alpha}_s GG \rangle}{12288\pi^4} \int_{x_i}^{1} dx\int_{y_i}^{1} dy\, (1-y)^2\,(u-\tilde{m}_y^2)\,\delta(s-\tilde{m}_x^2)\, ,
\end{eqnarray}

\begin{eqnarray}
\rho^{\bar{D}_s\Lambda_c}_{5}(s,u)
&=& -\frac{m_c^2\langle\bar{s}Gs\rangle}{1024\pi^4 T^4}\left(1 -\frac{m_s m_c}{3T^2}\right) \int_{y_i}^{1} dy\, y(1-y)^2\,(u-\tilde{m}_y^2)^2\,\delta(s-m_c^2) \nonumber\\
&& -\frac{m_c\langle\bar{q}Gq\rangle}{512\pi^4} \int_{x_i}^{1} dx\int_{y_i}^{1} dy\, \left[(1-x)m_c+xm_s\right]\,\delta(u-\tilde{m}_y^2) \nonumber\\
&& +\frac{\langle\bar{q}Gq\rangle}{1536\pi^4} \int_{x_i}^{1} dx\int_{y_i}^{1} dy\, y(1-y)\,(s-u)\,\delta(s-\tilde{m}_x^2) \nonumber\\
&& -\frac{(m_s+m_c) m_c\langle\bar{q}Gq\rangle}{6144\pi^4} \int_{x_i}^{1} dx\int_{y_i}^{1} dy\, (1-y)\,\delta(s-\tilde{m}_x^2) \nonumber\\
&& +\frac{m_c^2\langle\bar{q}Gq\rangle}{1536\pi^4} \int_{x_i}^{1} dx\int_{y_i}^{1} dy\, \frac{(1-x)(1-y)}{x}\,\delta(s-\tilde{m}_x^2) \nonumber\\
&& -\frac{\langle\bar{s}Gs\rangle}{2048\pi^4} \int_{y_i}^{1} dy\, (1-y)^2\,(u-\tilde{m}_y^2)\,\delta(s-m_c^2) \nonumber\\
&& +\frac{ m_c\langle\bar{q}Gq\rangle}{512\pi^4} \int_{x_i}^{1} dx\int_{y_i}^{1} dy\, \frac{(1-y)}{y}\,\left[(1-x)m_c+xm_s\right]\,\delta(u-\tilde{m}_y^2)\, ,
\end{eqnarray}

\begin{eqnarray}
\rho^{\bar{D}_s\Lambda_c}_{6}(s,u)
&=& -\frac{\langle\bar{q}q\rangle^2}{48\pi^2} \int_{x_i}^{1} dx\, \left[(1-x)m_c+xm_s\right]\,\delta(u-m_c^2) \nonumber\\
&& -\frac{m_c\langle\bar{q}q\rangle\langle\bar{s}s\rangle}{48\pi^2}\left(1-\frac{m_sm_c}{2T^2} \right) \int_{y_i}^{1} dy\, (1-y)\,\delta(s-m_c^2)\, ,
\end{eqnarray}

\begin{eqnarray}
\rho^{\bar{D}_s\Lambda_c}_{7}(s,u)
&=& -\frac{m_c^2\langle\bar{s}s\rangle\langle \tilde{\alpha}_s GG \rangle}{4608\pi^2 T^6}\left(1-\frac{m_sm_c}{2T^2} \right) \int_{y_i}^{1} dy\, y(1-y)^2\,(u-\tilde{m}_y^2)^2\,\delta(s-{m}_c^2) \nonumber\\
&& -\frac{m_c^3\langle\bar{q}q\rangle\langle \tilde{\alpha}_s GG \rangle}{1152\pi^2 T^4} \int_{x_i}^{1} dx\int_{y_i}^{1} dy\, \frac{(1-y)}{x^3}\,\left[(1-x)m_c+xm_s\right]\,\delta(s-\tilde{m}_x^2) \nonumber\\
&& +\frac{m_c^2\langle\bar{q}q\rangle\langle \tilde{\alpha}_s GG \rangle}{384\pi^2 T^2} \int_{x_i}^{1} dx\int_{y_i}^{1} dy\, \frac{(1-y)}{x^2}\,\delta(s-\tilde{m}_x^2) \nonumber\\
&& -\frac{m_s m_c\langle\bar{s}s\rangle\langle \tilde{\alpha}_s GG \rangle}{3072\pi^2 T^6} \int_{y_i}^{1} dy\, y(1-y)^2\,(u-\tilde{m}_y^2)^2\,\delta(s-{m}_c^2) \nonumber\\
&& -\frac{m_c^2\langle\bar{s}s\rangle\langle \tilde{\alpha}_s GG \rangle}{2304\pi^2}\left(1-\frac{m_sm_c}{2T^2} \right) \int_{y_i}^{1} dy\, \frac{(1-y)^2}{y^2}\,\delta(s-{m}_c^2)\,\delta(u-\tilde{m}_y^2) \nonumber\\
&& -\frac{ m_c^3\langle\bar{q}q\rangle\langle \tilde{\alpha}_s GG \rangle}{1152\pi^2 T^4} \int_{x_i}^{1} dx\int_{y_i}^{1} dy\, \frac{(1-y)}{y^3}\,\left[(1-x)m_c+xm_s\right]\,\delta(u-\tilde{m}_y^2) \nonumber\\
&& +\frac{ m_c\langle\bar{q}q\rangle\langle \tilde{\alpha}_s GG \rangle}{384\pi^2 T^2} \int_{x_i}^{1} dx\int_{y_i}^{1} dy\, \frac{(1-y)}{y^2}\,\left[(1-x)m_c+xm_s\right]\,\delta(u-\tilde{m}_y^2) \nonumber\\
&& +\frac{\langle\bar{s}s\rangle\langle \tilde{\alpha}_s GG \rangle}{768\pi^2} \left(1-\frac{m_sm_c}{2T^2} \right)\int_{y_i}^{1} dy\, y\,\delta(s-m_c^2)\nonumber\\
&& +\frac{\langle\bar{q}q\rangle\langle \tilde{\alpha}_s GG \rangle}{3072\pi^2} \int_{x_i}^{1} dx\int_{y_i}^{1} dy\, y\,(s-u)\,\delta(s-\tilde{m}_x^2)\,\delta(u-\tilde{m}_y^2) \nonumber\\
&& +\frac{m_s m_c\langle\bar{q}q\rangle\langle \tilde{\alpha}_s GG \rangle}{9216\pi^2} \int_{x_i}^{1} dx\int_{y_i}^{1} dy\, \frac{(2-5y)}{y}\,\delta(s-\tilde{m}_x^2)\,\delta(u-\tilde{m}_y^2) \nonumber\\
&& +\frac{\langle\bar{q}q\rangle\langle \tilde{\alpha}_s GG \rangle}{9216\pi^2} \int_{x_i}^{1} dx\int_{y_i}^{1} dy\, \frac{4-2y^2 + y - 2x(y + 2)  }{y}\,s\,\delta(s-\tilde{m}_x^2)\,\delta(u-\tilde{m}_y^2) \nonumber\\
&& -\frac{\langle\bar{s}s\rangle\langle \tilde{\alpha}_s GG \rangle}{9216\pi^2 T^2}\left(1+\frac{m_sm_c}{2T^2} \right) \int_{y_i}^{1} dy\, (1-y)^2\,(u-\tilde{m}_y^2)\,\delta(s-m_c^2) \nonumber\\
&& +\frac{m_c^4\langle\bar{s}s\rangle\langle \tilde{\alpha}_s GG \rangle}{9216\pi^2 T^8} \int_{y_i}^{1} dy\, y(1-y)^2\,(u-\tilde{m}_y^2)^2\,\delta(s-m_c^2) \nonumber\\
&& +\frac{ m_c\langle\bar{q}q\rangle\langle \tilde{\alpha}_s GG \rangle}{2304\pi^2 T^2} \int_{x_i}^{1} dx\, \left[(1-x)m_c+xm_s\right]\,\delta(u-m_c^2) \, ,
\end{eqnarray}

\begin{eqnarray}
\rho^{\bar{D}_s\Lambda_c}_{8}(s,u)
&=& \frac{m_c\langle\bar{s}s\rangle\langle\bar{q}Gq\rangle}{1536\pi^2} \left( 7-\frac{2m_sm_c}{T^2}\right)\int_{y_i}^{1} dy\,\delta(s-m_c^2)\,\delta(u-\tilde{m}_y^2) \nonumber\\
&& +\frac{\langle\bar{q}q\rangle\langle\bar{q}Gq\rangle}{96\pi^2 T^2} \left(1 + \frac{m_c^2}{T^2}\right) \int_{x_i}^{1} dx\, \left[(1-x)m_c+xm_s\right]\,\delta(u-m_c^2) \nonumber\\
&& +\frac{m_s\langle\bar{q}q\rangle\langle\bar{s}Gs\rangle}{192\pi^2 T^2}\left(1 + \frac{m_c^2}{T^2}-\frac{m_c^4}{3T^4}\right) \int_{y_i}^{1} dy\, (1-y)\,\delta(s-m_c^2) \nonumber\\
&& -\frac{m_s\langle\bar{s}s\rangle\langle\bar{q}Gq\rangle}{2304\pi^2 T^2} \int_{y_i}^{1} dy\, y(1-y)\,\left(2+\frac{u - m_c^2}{T^2} \right)\,\delta(s-m_c^2) \nonumber\\
&& -\frac{m_c\langle\bar{s}s\rangle\langle\bar{q}Gq\rangle}{4608\pi^2 T^2} \left(23-\frac{2m_sm_c}{T^2} \right)\int_{y_i}^{1} dy\, (1-y)\,\delta(s-m_c^2) \nonumber\\
&& -\frac{m_c\langle\bar{s}s\rangle\langle\bar{q}Gq\rangle}{384\pi^2} \left(1-\frac{m_sm_c}{2T^2}\right)\int_{y_i}^{1} dy\, \frac{1-y}{y}\,\delta(s-m_c^2)\,\delta(u-\tilde{m}_y^2) \, ,
\end{eqnarray}

\begin{eqnarray}
\rho^{\bar{D}_s\Lambda_c}_{9}(s,u)
&=& \frac{\langle\bar{q}q\rangle^2\langle\bar{s}s\rangle}{36}\left( 1-\frac{m_sm_c}{2T^2}\right)\,\delta(s-m_c^2)\,\delta(u-m_c^2) \, ,
\end{eqnarray}

\begin{eqnarray}\label{DsLambdac-10}
\rho^{\bar{D}_s\Lambda_c}_{10}(s,u)
&=& -\frac{m_c^4\langle\bar{q}Gq\rangle^2}{768\pi^2 T^8} \int_{x_i}^{1} dx\, \left[(1-x)m_c+xm_s\right]\,\delta(u-m_c^2) \nonumber\\
&& -\frac{m_c\langle\bar{q}Gq\rangle\langle\bar{s}Gs\rangle}{12288\pi^2 T^2} \left(15 + \frac{8m_c^2}{T^2} +\frac{8m_sm_c^3}{3T^4}\right) \int_{y_i}^{1} dy\,\delta(s-m_c^2)\,\delta(u-\tilde{m}_y^2) \nonumber\\
&& +\frac{m_s m_c^4\langle\bar{q}Gq\rangle\langle\bar{s}Gs\rangle}{2304\pi^2 T^6} \int_{y_i}^{1} dy\,\delta(s-m_c^2)\,\delta(u-\tilde{m}_y^2) \nonumber\\
&& +\frac{m_c^2\langle\bar{q}q\rangle^2\langle \tilde{\alpha}_s GG \rangle}{864 T^4} \int_{x_i}^{1} dx\, \frac{1}{x^3}\,\left[(1-x)m_c+xm_s\right]\,\delta(s-\tilde{m}_x^2)\,\delta(u-{m}_c^2) \nonumber\\
&& +\frac{m_c^3\langle\bar{q}q\rangle\langle\bar{s}s\rangle\langle \tilde{\alpha}_s GG \rangle}{864 T^6}\left( 1-\frac{m_sm_c}{2T^2}\right) \int_{y_i}^{1} dy\, (1-y)\,\delta(s-m_c^2) \nonumber\\
&& -\frac{m_c\langle\bar{q}q\rangle^2\langle \tilde{\alpha}_s GG \rangle}{3456 T^2} \int_{x_i}^{1} dx\, \frac{12+x-2x^2}{x^2}\,\delta(s-\tilde{m}_x^2)\,\delta(u-m_c^2) \nonumber\\
&& +\frac{m_s m_c^2\langle\bar{q}q\rangle\langle\bar{s}s\rangle\langle \tilde{\alpha}_s GG \rangle}{576 T^6}\left(1-\frac{m_c^2}{3T^2}\right) \int_{y_i}^{1} dy\, (1-y)\,\delta(s-m_c^2) \nonumber\\
&& +\frac{m_c^2\langle\bar{q}q\rangle^2\langle \tilde{\alpha}_s GG \rangle}{864 T^6}\left(1-\frac{m_c^2}{T^2} \right) \int_{x_i}^{1} dx\, \left[(1-x)m_c+xm_s\right]\,\delta(u-{m}_c^2) \nonumber\\
&& -\frac{m_c\langle\bar{q}q\rangle\langle\bar{s}s\rangle\langle \tilde{\alpha}_s GG \rangle}{288 T^2}\left(1-\frac{m_sm_c}{2T^2}\right) \int_{y_i}^{1} dy\, \frac{1-y}{y^2}\left(1-\frac{u}{3T^2} \right)\,\delta(s-m_c^2)\,\delta(u-\tilde{m}_y^2) \nonumber\\
&& +\frac{m_c\langle\bar{q}q\rangle\langle\bar{s}s\rangle\langle \tilde{\alpha}_s GG \rangle}{2304 T^2} \left( 1+\frac{m_sm_c}{2T^2}\right)\int_{y_i}^{1} dy\,\delta(s-m_c^2)\,\delta(u-\tilde{m}_y^2) \nonumber\\
&& -\frac{m_s\langle\bar{q}q\rangle\langle\bar{s}s\rangle\langle \tilde{\alpha}_s GG \rangle}{3456 T^2} \left(1-\frac{m_c^2}{T^2} \right)\int_{y_i}^{1} dy\, \frac{1-y}{y}\,\delta(s-m_c^2)\,\delta(u-\tilde{m}_y^2) \nonumber\\
&& +\frac{m_s\langle\bar{q}q\rangle^2\langle \tilde{\alpha}_s GG \rangle}{3456 T^2} \int_{x_i}^{1} dx\,\delta(s-\tilde{m}_x^2)\,\delta(u-m_c^2) \nonumber\\
&& -\frac{m_s\langle\bar{q}q\rangle\langle\bar{s}s\rangle\langle \tilde{\alpha}_s GG \rangle}{13824 T^2} \int_{y_i}^{1} dy\, (2+y)\,\left(2 +\frac{u- m_c^2}{T^2} \right)\,\delta(s-m_c^2)\,\delta(u-\tilde{m}_y^2) \nonumber\\
&& -\frac{m_c\langle\bar{q}Gq\rangle\langle\bar{s}Gs\rangle}{2048\pi^2 T^2}\,\delta(s-m_c^2)\,\delta(u-{m}_c^2) \nonumber\\
&& +\frac{ m_c^2\langle\bar{q}Gq\rangle\langle\bar{s}Gs\rangle}{55296\pi^2 T^6}\left[4m_s\left(1 - \frac{m_c^2}{T^2}\right)-3m_c\right] \int_{y_i}^{1} dy\, (1-y)\,\delta(s-m_c^2) \nonumber\\
&& +\frac{m_sm_c^2\langle\bar{q}Gq\rangle\langle\bar{s}Gs\rangle}{13824\pi^2 T^6} \int_{y_i}^{1} dy\, y(1-y)\,\left(4 +\frac{u-m_c^2}{T^2} \right)\,\delta(s-m_c^2) \nonumber\\
&& -\frac{m_c\langle\bar{q}Gq\rangle\langle\bar{s}Gs\rangle}{1536\pi^2T^2}\left(2 -\frac{m_c^2}{T^2}+\frac{m_sm_c^3}{3T^4} \right) \int_{y_i}^{1} dy\, \frac{1-y}{y}\,\delta(s-m_c^2)\,\delta(u-\tilde{m}_y^2) \nonumber\\
&& +\frac{\langle\bar{q}Gq\rangle^2}{27648\pi^2 T^2} \int_{x_i}^{1} dx\, \left(\frac{19-2x}{x}m_c+2m_s\right)\,\delta(s-\tilde{m}_x^2)\,\delta(u-m_c^2) \nonumber\\
&& -\frac{m_c\langle\bar{q}q\rangle\langle\bar{s}s\rangle\langle \tilde{\alpha}_s GG \rangle}{1728 T^2}\left(1-\frac{m_c^2}{2T^2} \right)\,\delta(s-m_c^2)\,\delta(u-m_c^2)\, ,
\end{eqnarray}
where $\bar{\alpha}_s=\frac{\alpha_s}{\pi}$, $\langle \bar{q}Gq\rangle= \langle\bar{q}g_s\sigma Gq\rangle$, $\langle \bar{s}Gs\rangle= \langle\bar{s}g_s\sigma Gs\rangle$, $x_{f}=\frac{1+\sqrt{1-4m_{c}^{2}/s}}{2}$, $x_{i}=\frac{1-\sqrt{1-4m_{c}^{2}/s}}{2}$,   $\widetilde{m}_{c}^{2}=\frac{m_{c}^{2}}{x(1-x)}$ in Eqs.\eqref{etacLambda-0}-\eqref{JpsiLambda-10},
$x_i=\frac{m_c^2}{s}$, $y_i=\frac{m_c^2}{u}$, $\widetilde{m}_x^2=\frac{m_c^2}{x}$, $\widetilde{m}_y^2=\frac{m_c^2}{y}$ in Eqs.\eqref{DsLambdac-0}-\eqref{DsLambdac-10}, $\int_{x_{i}}^{x_{f}}dx\rightarrow\int_{0}^{1} dx$, $\int_{x_{i}}^{1}dx\rightarrow\int_{0}^{1} dx$ and $\int_{y_{i}}^{1}dy\rightarrow\int_{0}^{1} dy$, when the $\delta$ functions $\delta(s-\widetilde{m}_{c}^{2})$, $\delta(s-\widetilde{m}_{x}^{2})$ and $\delta(u-\widetilde{m}_{y}^{2})$ appear.

\section*{Acknowledgements}
This  work is supported by National Natural Science Foundation, Grant Number  12575083.


\begin{thebibliography}{99}

\bibitem{LHCb-Pcs4338-2023}  R. Aaij et al,  Phys. Rev. Lett. {\bf 131} (2023)  031901.

\bibitem{PDG}  F. Takahashi et al, Int. J. Mod. Phys. {\bf A41} (2026) 2630011.

\bibitem{WangZG-review} Z. G. Wang,  Front. Phys. {\bf 21} (2026) 016300.


\bibitem{Pcc4338-mole-WangFL-PLB-2022} F. L. Wang and  X. Liu, Phys. Lett.
{\bf B835} (2022) 137583.


\bibitem{Pcc4338-mole-Feijoo-PLB-2023} A. Feijoo, W. F. Wang, C. W. Xiao, J. J. Wu, E. Oset,  J. Nieves and B. S. Zou, Phys. Lett. {\bf B839} (2023) 137760.


\bibitem{Pcc4338-mole-Ozdem-PLB-2023} U. Ozdem, Phys. Lett. {\bf B836} (2023) 137635.

\bibitem{Pcc4338-mole-MengL-PRD-2023} L. Meng, B. Wang and S. L. Zhu,
 Phys. Rev. {\bf D107} (2023)  014005.

\bibitem{Pcc4338-mole-Ortega-PLB-2023} P. G. Ortega, D. R. Entem and F. Fernandez, Phys. Lett. {\bf B838} (2023) 137747.

\bibitem{Pcc4338-mole-YanMJ-PRD-2023} M. J. Yan, F. Z. Peng, M. Sanchez and M. P. Valderrama, Phys. Rev. {\bf D107} (2023) 074025.

\bibitem{Pcc4338-mole-Nakamura-PRD-2023}  S. X. Nakamura and J. J. Wu,
Phys. Rev. {\bf D108} (2023)  L011501.

\bibitem{Pcc4338-mole-HeJ-PRD-2023} J. T. Zhu, S. Y. Kong and J. He, Phys. Rev. {\bf D107} (2023)  034029.


\bibitem{Pcc4338-mole-LiHS-PRD-2024} H. S. Li, F. Guo, Y. D. Lei and F. Gao, Phys. Rev. {\bf D109} (2024)  094027.

\bibitem{Pcc4338-mole-LiHS-PRD-2024-2} H. S. Li,  Phys. Rev. {\bf D109} (2024)  114039.

\bibitem{Pcc4338-mole-WangB-PRD-2024} B. Wang, K. Chen, L. Meng and S. L. Zhu,  Phys. Rev. {\bf D109} (2024) 074035.

\bibitem{Pcc4338-mole-PingJL-Symmetry-2024}
G. Yang, J. Ping and J. Segovia, Symmetry {\bf 16} (2024)  354.

\bibitem{Pcs4338-mole-WangXW-CPC-2023}
X. W. Wang and Z. G. Wang, Chin. Phys. {\bf C47} (2023)  013109.



\bibitem{Pcs4338-mole-decay-WangXW-PRD} X. W. Wang and Z. G. Wang,  Phys. Rev.
{\bf D110} (2024)  014008.

\bibitem{Pcs4338-mole-decay-Azizi-PRD-2023}
K. Azizi, Y. Sarac and H. Sundu, Phys. Rev. {\bf D108} (2023)  074010.

\bibitem{Pcs4338-penta-MM-Ozdem-PRD-2026} U. Ozdem, Phys. Rev. {\bf D114} (2026)  014050.

\bibitem{Pcs4338-Braaten-BO} F. Alasiri, E. Braaten and R. Bruschini,
arXiv: 2610.06009 [hep-ph].

\bibitem{Pcc4338-Product-Clymton-PRD-2025} S. Clymton, H. C. Kim and T. Mart,
 Phys. Rev. {\bf D112} (2025)  014041.

\bibitem{Pcs4338-Product-HeJ-EPJC-2025}  S. Chen, J. T. Zhu, S. Y. Kong and J. He, Eur. Phys. J. {\bf C85} (2025)  492.

\bibitem{Pcs4338-Product-HsiaoYK-PRD-2025}
Y. K. Hsiao, S. T. Cai and Y. L. Wang, Phys. Rev. {\bf D111} (2025) 076020.

\bibitem{Pcs4338-Product-ChenDY-PRD-2024}
 Q. Wu and D. Y. Chen, Phys. Rev. {\bf D109} (2024)  094003.


\bibitem{WangZG-penta-Jpsi-p}  Z. G. Wang, Eur. Phys. J. {\bf C86}  (2026) 209.

\bibitem{WangZG-penta-IJMPA-2020} Z. G. Wang,  Int. J. Mod. Phys. {\bf A35}
    (2020)  2050003.

\bibitem{WangZG-penta-Jpsi-Lamba}  Z. G. Wang and Q. Xin, Eur. Phys. J. {\bf C86} (2026) 472.


\bibitem{WangZG-penta-Jpsi-Xi}  Z. G. Wang and Y. Liu, Nucl. Phys. {\bf B1027} (2026) 117456.

\bibitem{WangZG-penta-Jpsi-Sigma}  Z. G. Wang and Y. Liu, Front. Phys. {\bf 22} (2027) 016201.

\bibitem{WangZG-penta-Jpsi-XiV}  Z. G. Wang, arXiv:2606.28095 [hep-ph].

\bibitem{WangZG-penta-Jpsi-Omega}  Z. G. Wang, Nucl. Phys. {\bf B1032} (2026) 117678.

\bibitem{WangZG-penta-Jpsi-SigmaV}  Z. G. Wang and Y. Liu, arXiv:2607.17492 [hep-ph].

\bibitem{QCDSR-Narison-PRD-2007}  R. D. Matheus, S. Narison, M. Nielsen and J. M. Richard, Phys. Rev. {\bf D75} (2007) 014005.

\bibitem{QCDSR-Nielsen-PRT-2010}
 M. Nielsen, F. S. Navarra and S. H. Lee, Phys. Rept. {\bf 497} (2010) 41.


\bibitem{QCDSR-ChenZhu-PRD-2011}
 W. Chen and S. L. Zhu, Phys. Rev. {\bf D83} (2011) 034010.

 \bibitem{QCDSR-Nielsen-JPG-2019}
 R. M. Albuquerque, J. M. Dias, K. P. Khemchandani, A. M. Torres, F. S. Navarra, M. Nielsen and C. M. Zanetti, J. Phys. {\bf G46} (2019)  093002.

\bibitem{QCDSR-ChenW-PRD-2013}
 M. L. Du, W. Chen, X. L. Chen and S. L. Zhu, Phys. Rev. {\bf D87} (2013)  014003.

\bibitem{QCDSR-ZhangJR-PLB-2011}
 J. R. Zhang, M. Zhong and M. Q. Huang, Phys. Lett. {\bf B704} (2011) 312.

\bibitem{QCDSR-ZhangJR-PRD-2013}
J. R. Zhang, Phys. Rev. {\bf D87} (2013)  116004.

\bibitem{QCDSR-QiaoCF-NPB-2021}
B. D. Wan and C. F. Qiao, Nucl. Phys. {\bf B968} (2021) 115450.

\bibitem{QCDSR-QiaoCF-PRD-2020}
L. Tang, B. D. Wan, K. Maltman and C. F. Qiao, Phys. Rev. {\bf D101} (2020)  094032.


\bibitem{WangZG-HuangT-PRD-2014}
Z. G. Wang and T. Huang, Phys. Rev. {\bf D89} (2014)  054019.

\bibitem{WangZG-formula-EPJC} Z. G. Wang, Eur. Phys. J. {\bf C74} (2014)  2874.

\bibitem{Wang1508-EPJC-1} Z. G. Wang, Eur. Phys. J. {\bf C76} (2016) 70.


\bibitem{Decay-Nielsen-PRD-2013} J. M. Dias, F. S. Navarra, M. Nielsen and C. M. Zanetti, Phys. Rev. {\bf D88} (2013)  016004.

\bibitem{Decay-WangZG-NPA-2014}  Z. G. Wang and  T. Huang, Nucl. Phys. {\bf A930} (2014) 63.

\bibitem{Decay-ChenHX-EPJC-2015}
W. Chen, T. G. Steele, H. X. Chen and S. L. Zhu, Eur. Phys. J. {\bf C75} (2015)  358.


\bibitem{Decay-Azizi-PRD-2016}
S. S. Agaev, K. Azizi and H. Sundu, Phys. Rev. {\bf D93} (2016)  074002.

\bibitem{Duality-WangZG-EPJC-2016} Z. G. Wang, Eur. Phys. J. {\bf C76} (2016)  279.

\bibitem{Duality-WangZG-ZhangJX-EPJC-2018} Z. G. Wang and J. X. Zhang, Eur. Phys. J. {\bf C78} (2018)  14.


\bibitem{Decay-Azizi-EPJC-2019}
H. Sundu, S. S. Agaev and K. Azizi, Eur. Phys. J. {\bf C79} (2019)  215.

\bibitem{Duality-WangZG-EPJC-2019} Z. G. Wang, Eur. Phys. J. {\bf C79} (2019)  184.

\bibitem{Duality-Penta-mole-WangZG-CPC-2020}
 Z. G. Wang and X. Wang, Chin. Phys. {\bf C44} (2020) 103102.


\bibitem{Duality-Penta-mole-LiuYL-PRD-2020}
Y. J. Xu, C. Y. Cui, Y. L. Liu and M. Q. Huang, Phys. Rev. {\bf D102} (2020)  034028.


\bibitem{Duality-Penta-WangZG-CPC-2021}
Z. G. Wang, H. J. Wang and Q. Xin, Chin. Phys. {\bf C45} (2021) 063104.

\bibitem{Decay-Azizi-JPG-2021}
S. S. Agaev, K. Azizi and H. Sundu, J. Phys. {\bf G48} (2021)  085012.

\bibitem{Duality-WangZG-CPC-2022}
Z. G. Wang, Chin. Phys. {\bf C46} (2022)  123106.

\bibitem{Duality-WangZG-CPC-2022-2}
Z. G. Wang, Chin. Phys. {\bf C46} (2022)  103106.

\bibitem{Decay-Azizi-EPJC-2023}
H. Sundu, S. S. Agaev and K. Azizi, Eur. Phys. J. {\bf C83} (2023)  198.

\bibitem{Decay-SunH-NPB-2023}
Y. Xie, D. He, X. Luo and H. Sun, Nucl. Phys. {\bf B987} (2023) 116113.

\bibitem{Duality-WangZG-PRD-2024}
Z. G. Wang, Phys. Rev. {\bf D109} (2024)  014017.


\bibitem{Duality-WangZG-AAPPS-2024}
Z. G. Wang and X. S. Yang, AAPPS Bull. {\bf 34} (2024) 5.

\bibitem{Duality-YangXS-CPC-2025} X. S. Yang and Z. G. Wang, Chin. Phys. {\bf C49} (2025)  063108.

\bibitem{Duality-WangXW-CPC-2024} X. W. Wang and Z. G. Wang, Chin. Phys.
{\bf C48} (2024)  053102.


\bibitem{Duality-WangZG-NPB-2024}
Z. G. Wang, Nucl. Phys. {\bf B1005} (2024) 116580.


\bibitem{Decay-Azizi-NPB-2026}
S. S. Agaev, K. Azizi and H. Sundu, Nucl. Phys. {\bf B1031} (2026) 117652.

\bibitem{Duality-YangXS-CPC-2026}
X. S. Yang and Z. G. Wang, Chin. Phys. {\bf C50} (2026)  093104.

\bibitem{SVZ79}  M. A. Shifman, A. I. Vainshtein and V. I. Zakharov, Nucl. Phys. {\bf B147} (1979) 385, 448.

\bibitem{PRT85} L. J. Reinders, H. Rubinstein and S. Yazaki, Phys. Rept. {\bf 127} (1985) 1.


\bibitem{WangZG-Landau-PRD-2020} Z. G. Wang, Phys. Rev. {\bf D101} (2020) 074011.




\bibitem{Becirevic} D. Becirevic, G. Duplancic, B. Klajn, B. Melic and F. Sanfilippo,  Nucl. Phys. {\bf B883} (2014) 306.

\bibitem{WangZG-decay-cnstant-D}  Z. G. Wang, Eur. Phys. J. {\bf C75} (2015) 427.

\bibitem{WangZG-Lambdac} Z. G. Wang, Eur. Phys. J. {\bf C68} (2010) 479.

\bibitem{WangZG-Lambdac-2S} Z. G. Wang and  H. J. Wang,  Chin. Phys. {\bf C45} (2021)  013109.

\bibitem{ColangeloReview} P. Colangelo and A. Khodjamirian, hep-ph/0010175.

\bibitem{Narison-mix} S. Narison and R. Tarrach, Phys. Lett. {\bf 125 B} (1983) 217.
\bibitem{ZhaoZX-gamma} Z. X. Zhao, Y. J. Shi and Z. P. Xing,
arxiv: 2104.06209 [hep-ph].

\bibitem{Latt-Decay-fDs} A. Bazavov et al, Phys. Rev. {\bf D98} (2018) 074512.

\end{thebibliography}
\end{document}